\documentclass[twocolumn]{aastex63}
\usepackage{bm}
\usepackage{ulem}
\usepackage{footmisc}

\received{October 15, 2020}
\revised{January 29, 2021}
\accepted{February 9, 2021}
\submitjournal{the Astronomical Journal}

\shorttitle{Searching for Low-mass Population III Stars}
\shortauthors{Chandra \& Schlaufman}

\begin{document}

\title{Searching for Low-mass Population III Stars Disguised as White
Dwarfs}

\correspondingauthor{Vedant Chandra}
\email{vchandra@jhu.edu}

\author[0000-0002-0572-8012]{Vedant Chandra}
\affiliation{Department of Physics and Astronomy \\
Johns Hopkins University \\
3400 N Charles St \\
Baltimore, MD 21218, USA}

\author[0000-0001-5761-6779]{Kevin C. Schlaufman}
\affiliation{Department of Physics and Astronomy \\
Johns Hopkins University \\
3400 N Charles St \\
Baltimore, MD 21218, USA}

\begin{abstract}

\noindent
It is uncertain whether or not low-mass Population III stars ever existed.
While limits on the number density of Population III stars with $M_{\ast}
\approx 0.8~M_{\odot}$ have been derived using Sloan Digital Sky Survey
(SDSS) data, little is known about the occurrence of Population III
stars at lower masses.  In the absence of reliable parallaxes, the
spectra of metal-poor main sequence (MPMS) stars with $M_{\ast} \lesssim
0.8~M_{\odot}$ can easily be confused with cool white dwarfs.  To resolve
this ambiguity, we present a classifier that differentiates between
MPMS stars and white dwarfs based on photometry and/or spectroscopy
without the use of parallax information.  We build and train our
classifier using state-of-the-art theoretical spectra and evaluate it on
existing SDSS-based classifications for objects with reliable Gaia DR2
parallaxes.  We then apply our classifier to a large catalog of objects
with SDSS photometry and spectroscopy to search for MPMS candidates.
We discover several previously unknown candidate extremely metal-poor
(EMP) stars and recover numerous confirmed EMP stars already in the
literature.  We conclude that archival SDSS spectroscopy has already been
exhaustively searched for EMP stars.  We predict that the lowest-mass
primordial-composition stars will have redder optical-to-infrared colors
than cool white dwarfs at constant effective temperature due to surface
gravity-dependent collision-induced absorption from molecular hydrogen.
We suggest that the application of our classifier to data produced by
next-generation spectroscopic surveys will set stronger constraints on
the number density of low-mass Population III stars in the Milky Way.

\end{abstract}

\keywords{Chemically peculiar stars (226) --- Low mass stars (2050) ---
Population II stars (1284) --- Population III stars (1285) ---
Sky surveys (1464) --- White dwarf stars (1799)}

\section{Introduction}\label{sec:intro}

The first generation of stars formed in the Universe was made of
only the stable products of Big Bang nucleosynthesis: hydrogen,
helium, and a tiny amount of lithium.  These Population III stars
are predicted to start forming around 100 Myr after the Big Bang
\citep[e.g.,][]{Bromm2013,Glover2013,Greif2015}.  The earliest Population
III star formation calculations suggested that inefficient cooling would
require large Jeans masses and therefore that Population III stars would
form with a characteristic stellar mass $M_{\ast} \sim 100~M_{\odot}$
\citep[e.g.,][]{Silk1983,Tegmark1997,Bromm1999,Bromm2002,Abel2000,Abel2002}.
However, more recent simulations have shown that
fragmentation in the accretion disks around massive Population
III protostars could potentially form pristine stars at much lower masses
\citep[e.g.,][]{Stacy2010,Stacy2012,Stacy2016,Clark2011a,Clark2011b,Greif2011,Greif2012,Stacy2013,Stacy2014,Dopcke2013,Riaz2018,Wollenberg2020}.
While it is theoretically uncertain if these fragments survive or
merge with the more massive protostar growing at the center of their
parent accretion disk \citep[e.g.,][]{Hirano2017}, there is at least
circumstantial observational evidence to suggest that they might
survive \citep{Schlaufman2018}.  If these fragments do avoid merging,
then low-mass Population III stars might persist to the present day in
the local Universe.

While no Population III star has been directly observed to
date, according to the latest Stellar Abundances for
Galactic Archaeology (SAGA) database observational searches
have instead found more than 500 extremely
metal-poor (EMP) stars with metallicity $[\text{Fe/H}] \lesssim
-3$ \citep{Suda2008,Suda2011,Suda2017,Yamada2013}.

The chemical abundances of these extreme Population II stars
can be used to infer the properties of Population III stars
\citep[e.g.,][]{Hartwig2015a,Placco2016,Fraser2017, Ishigaki2018,
Hansen2020}, as well as the early chemical evolution of the Milky Way
\citep[e.g.,][]{2005ARA&A..43..531B,2015ARA&A..53..631F, Kobayashi2020}.

Because stars with bright apparent magnitudes are more
easily studied than stars with faint apparent magnitudes, aside
from early examples found in studies of high proper motion stars
\citep[e.g.,][]{1991AJ....101.1835R,1991AJ....101.1865R,1991AJ....102..303R}
the vast majority of metal-poor stars studied in detail to date are at
least as luminous as the main sequence turnoff \citep{2013ApJ...762...26Y,
2013ApJ...778...56C,2014AJ....147..136R}.  Of the 8419 unique stars in
the SAGA database with effective temperature $T_{\text{eff}}$ and surface
gravity $\log{g}$ inferences, only 199 or 2.3\% have $T_{\text{eff}}
\lesssim 6000$ K and $\log{g} \gtrsim 4.6$ indicative of metal-poor main
sequence (MPMS) stars.  As a result, only a small fraction of the Milky
Way's metal-poor stellar population has been characterized to date.
Indeed, assuming the PARSEC version 1.2S  evolutionary tracks
\citep{Bressan2012,Chen2014,Chen2015} and a \cite{Kroupa2001,Kroupa2002a}
initial mass function for a 10 Gyr stellar population, the luminosity
function for stars with $[\text{Fe/H}] \approx -2.2$ indicates that
for every metal-poor turnoff star with $T_{\text{eff}} \gtrsim 6000$
K and $3.8 \lesssim \log{g} \lesssim 4.6$ there are about 20 MPMS stars
with $T_{\text{eff}} \lesssim 6000$ K and $\log{g} \gtrsim 4.6$ in the
same volume.

One
significant obstacle to the identification of MPMS stars is that
the optical spectra of metal-poor (and metal-free) main sequence
stars are virtually indistinguishable from the optical spectra of
more numerous cool hydrogen-atmosphere white dwarfs (WD).  While a
significant fraction of white dwarf photospheres are externally
seeded with metals by ongoing accretion from circumstellar disks
\citep[e.g.,][]{Gaensicke2012,Koester2014}, the high surface gravities
of white dwarfs cause metals to rapidly sink out of their photospheres
via gravitational settling \citep{Schatzman1948}.  The spectra of most white dwarf photospheres
are therefore very similar to those of metal-free stars, featuring
strong Balmer lines and possibly other absorption features due to
molecular hydrogen.  While in the absence of precise parallaxes
it is sometimes possible to separate MPMS stars from white dwarfs
using reduced proper motion as an analog for absolute magnitude
\citep[e.g.,][]{Luyten1922}, reduced proper motions are ambiguous for
individual stars and can be confounded if the samples of interest belong
to distinct kinematic populations.

The traditional method to distinguish MPMS stars from white dwarfs has
been to visually identify MPMS stars on the basis of their narrower Balmer
absorption lines at a given photometric color \citep{Kepler2019a}.  This
process is prone to human error and very difficult for large data sets.
These problems have historically been aggravated by imperfect models for
the atmospheres of cool white dwarfs \citep{Kepler2019a}.  Recent advances
have improved the fidelity of atmosphere models for white dwarfs with
$T_{\text{eff}} \lesssim 5000$ K \citep{Blouin2018,Blouin2018a},
facilitating a theoretical comparison between MPMS star and white
dwarf spectra at the coolest temperatures where spectroscopic data are
currently lacking.

Tens of thousands of spectra with weak or non-existent
metal lines produced by either cool white dwarfs or
metal-poor/metal-free main sequence stars will be collected
by ongoing and next-generation spectroscopic facilities/surveys
like the Large Sky Area Multi-Object Fibre Spectroscopic Telescope
\citep[LAMOST;][]{2012RAA....12.1197C}, the Dark Energy Spectroscopic
Instrument \citep[DESI;][]{2016arXiv161100036D}, the fifth phase
of the Sloan Digital Sky Survey \citep[SDSS-V;][]{Kollmeier2017},
and the WHT Enhanced Area Velocity Explorer \citep[WEAVE;][]{WEAVE}.
The volume of data produced by these ongoing and future surveys will
make it impossible to rely on the traditional human classification of
MPMS stars and white dwarfs.  This implies that there is a need for an
automated way to differentiate between MPMS stars and white dwarfs.

In this paper, we develop an automated framework to identify MPMS
stars in large spectroscopic surveys with a focus on the problem of
differentiating MPMS stars from spectroscopically similar white dwarfs.
We describe in Section \ref{sec:data} state-of-the-art theoretical
spectra for both MPMS stars and white dwarfs.  We then introduce the
Sloan Digital Sky Survey (SDSS) photometric and spectroscopic data for
MPMS stars and white dwarfs securely classified using Gaia Data Release 2
(DR2) parallaxes that we use to validate and test our methods.  We outline
our classification algorithm in Section \ref{sec:methods}, including our
extraction of features from photometric and spectroscopic observables
and the validation of our classifier.  In Section \ref{sec:results} we
perform a search for candidate extremely metal-poor stars in a sample of
possible MPMS stars with SDSS photometry and spectroscopy.  We review
our results and their implications in Section \ref{sec:discussion}. We
conclude by summarizing our findings in Section \ref{sec:conclusion}.

\section{Data}\label{sec:data}

Our goal is to differentiate MPMS stars from white dwarfs on the
basis of photometry and spectroscopy alone.  While there is enough
observational data for MPMS stars and white dwarfs with $T_{\text{eff}}
\gtrsim 6000\text{~K}$, the relative faintness of MPMS stars and white
dwarfs with $T_{\text{eff}} \lesssim 6000\text{~K}$ has resulted in few
examples in the SDSS spectroscopic archive.  Even though these cool MPMS
stars and white dwarfs were largely out of reach of the Sloan Foundation
2.5\,m Telescope, spectroscopic surveys using 4\,m class telescopes like
DESI will target many cool MPMS stars and white dwarfs.  It is therefore
necessary for us to use both theoretical spectra for cool MPMS stars
and white dwarfs and empirical spectra for warmer MPMS stars and white
dwarfs to train and validate our classifier.  We describe those data in
the following two subsections.

\subsection{Theoretical Spectra}\label{sec:theorydata}

\begin{figure}
    \centering
    \includegraphics[width=\columnwidth]{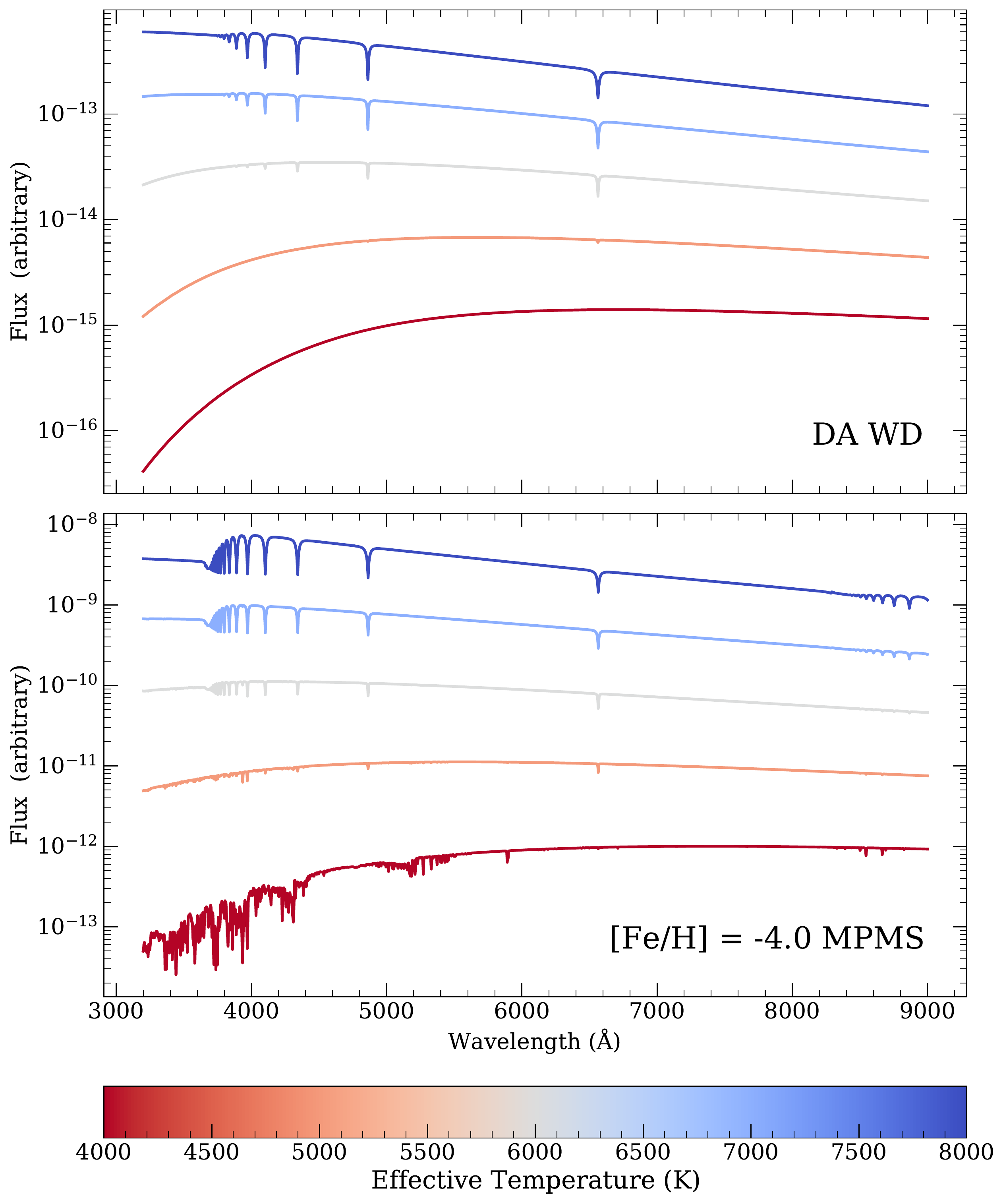}
    \caption{Sample theoretical white dwarf and MPMS star spectra from
    the \cite{Blouin2018} and PHOENIX \citep{Husser2013} libraries,
    respectively.  We plot spectra spanning the range $4000\text{~K} \leq
    T_{\text{eff}} \leq 8000\text{~K}$ assuming $\log{g} = 8$ and $\log{g}
    = 4.5$ for the white dwarf and MPMS star spectra, respectively.
    For the MPMS star spectra, we assume $\text{[Fe/H]} = -4.0$ with
    scaled solar abundances.  We offset the fluxes vertically for visual
    clarity.  The optical spectrum of a very low-mass Population III star
    would be spectroscopically indistinguishable from the $T_{\text{eff}}
    = 4000\text{~K}$ white dwarf in the top panel.\label{fig:overview}}
\end{figure}

We first construct grids of theoretical spectra for white dwarfs and
MPMS stars using state-of-the-art models.  We consider temperatures in
the range $4000\text{~K} \leq T_{\text{eff}} \leq 8000\text{~K}$ in 40 K
steps.  For each temperature, we generate white dwarf spectra with $7 \leq
\log{g} \leq 9$ and MPMS star spectra with $3.5 \leq \log{g} \leq 5.5$
in steps of 0.5 dex.  This procedure results in grids of 505 theoretical
spectra that we use to train our classifier.  For the white dwarfs,
our grid of theoretical spectra is described in  \citet{Blouin2018,Blouin2018a}.
These spectra were computed assuming
pure-hydrogen atmospheres, no magnetic fields, and metal-free
photospheres.  \citet{Blouin2018,Blouin2018a,Blouin2019} describe the
improved equation of state and radiative opacities featured in this new
generation of models that are especially important for cool white dwarfs
with $T_{\text{eff}} \lesssim 5000\text{~K}$.  For MPMS stars, we use
theoretical spectra computed with the PHOENIX code \citep{Husser2013}.
Those models cover a broad range of temperatures and surface gravities and
include metallicities $-4.0 \leq \text{[Fe/H]} \leq +1.0$ and $\alpha$
abundances $-0.2 \leq \text{[$\alpha$/Fe]} \leq +1.2$.  In our theoretical
grid of MPMS star spectra, we use the $\text{[Fe/H]} = -4.0$ spectra
with solar $\alpha$ abundances. The spectroscopic classifier we
describe in Section \ref{sec:methods} relies on Balmer lines, so the
assumed metallicity has only a negligible impact on our analyses.

For both the MPMS and WD grids of theoretical spectra spanning
$T_{\rm{eff}}$ and $\log{g}$, we tri-linearly interpolate the logarithm
of the fluxes with respect to effective temperature, surface gravity,
and the logarithm of wavelength.  We convolve all theoretical spectra
to match an instrumental resolution of $1.5$ \AA, comparable to the
resolution of large-scale spectroscopic surveys like the SDSS, LAMOST,
and DESI.  We illustrate some sample theoretical spectra in Figure
\ref{fig:overview}.

\subsection{SDSS Data}\label{sec:empiricaldata}

Our empirical data for warmer MPMS stars and white dwarfs come mostly
from SDSS DR16 \citep{Ahumada2020}.  They were collected during the first
four phases of the SDSS \citep{York2000,Eisenstein2011,Blanton2017},
including its Sloan Extension for Galactic Understanding and Exploration
(SEGUE), Baryon Oscillation Spectroscopic Survey (BOSS), and extended BOSS
(eBOSS) programs \citep{Yanny2009,Dawson2013,Dawson2016}.  The data were
collected using the Sloan Foundation 2.5\,m Telescope and its imager
and optical spectrographs \citep{Gunn1998,Gunn2006,Doi2010,Smee2013}
then placed on the SDSS photometric system using the methods
described in \citet{Fukugita1996}, \citet{Smith2002}, and
\citet{Padmanabhan2008}.  We also make use of Gaia DR2 astrometry
\citep{newGaia2016,newGaia2018,Salgado2017,Arenou2018,Lindegren2018,Luri2018,Marrese2019}.
We de-redden the SDSS \textit{ugriz} magnitudes using the \texttt{mwdust}
utility \citep{Bovy2016} and the combined dust maps of \cite{Drimmel2003},
\cite{Marshall2006}, and \cite{Green2019}.

We focus on the MPMS star and white dwarf classifications provided
by \cite{Kepler2019a}.  Those authors examined 500,000 SDSS spectra
plausibly produced by white dwarfs.  They provide classifications for
37,053 spectra based on 11 criteria, of which 15,716  were classified as
DA white dwarfs and 15,855 were classified as subdwarf A or sdA stars
(i.e., likely MPMS stars).  Among the white dwarfs, over 78\% were
classified as DA white dwarfs with hydrogen-rich atmospheres on the basis
of broad Balmer absorption lines with no other strong spectral features.
\cite{Kepler2019a} visually differentiated these DA white dwarfs from
subdwarf A stars based on the fact that DA white dwarfs have broader
Balmer lines at a given photometric color due to their high surface
gravities and increased pressure broadening.  Together these two
classes comprise more than 85\% of the classifications provided in
\cite{Kepler2019a}.

We use in our analysis only those white dwarfs or MPMS stars classified
by \cite{Kepler2019a} as members of their ``DA'', ``sdA'', or ``sdA/F''
classes. The class ``DA'' designates hydrogen-atmosphere
white dwarfs, the class ``sdA'' refers to subdwarf A stars, and the
class ``sdA/F'' indicates ambiguous subdwarf A or F stars.  This
limits our sample to 14,522 spectra with prominent Balmer lines
and no strong metal features.  It also removes other white dwarf
spectral types like helium-rich DB white dwarfs.  While most of
the stars classified as ``sdA'' or ``sdA/F'' stars are likely MPMS
stars \citep[e.g.,][]{Brown2017,Pelisoli2018,Pelisoli2018b}, a very
small fraction could also be extremely low-mass (ELM) white dwarfs
\citep[e.g.,][]{Kosakowski2020}.

\begin{figure}
    \centering
    \includegraphics[width=\columnwidth]{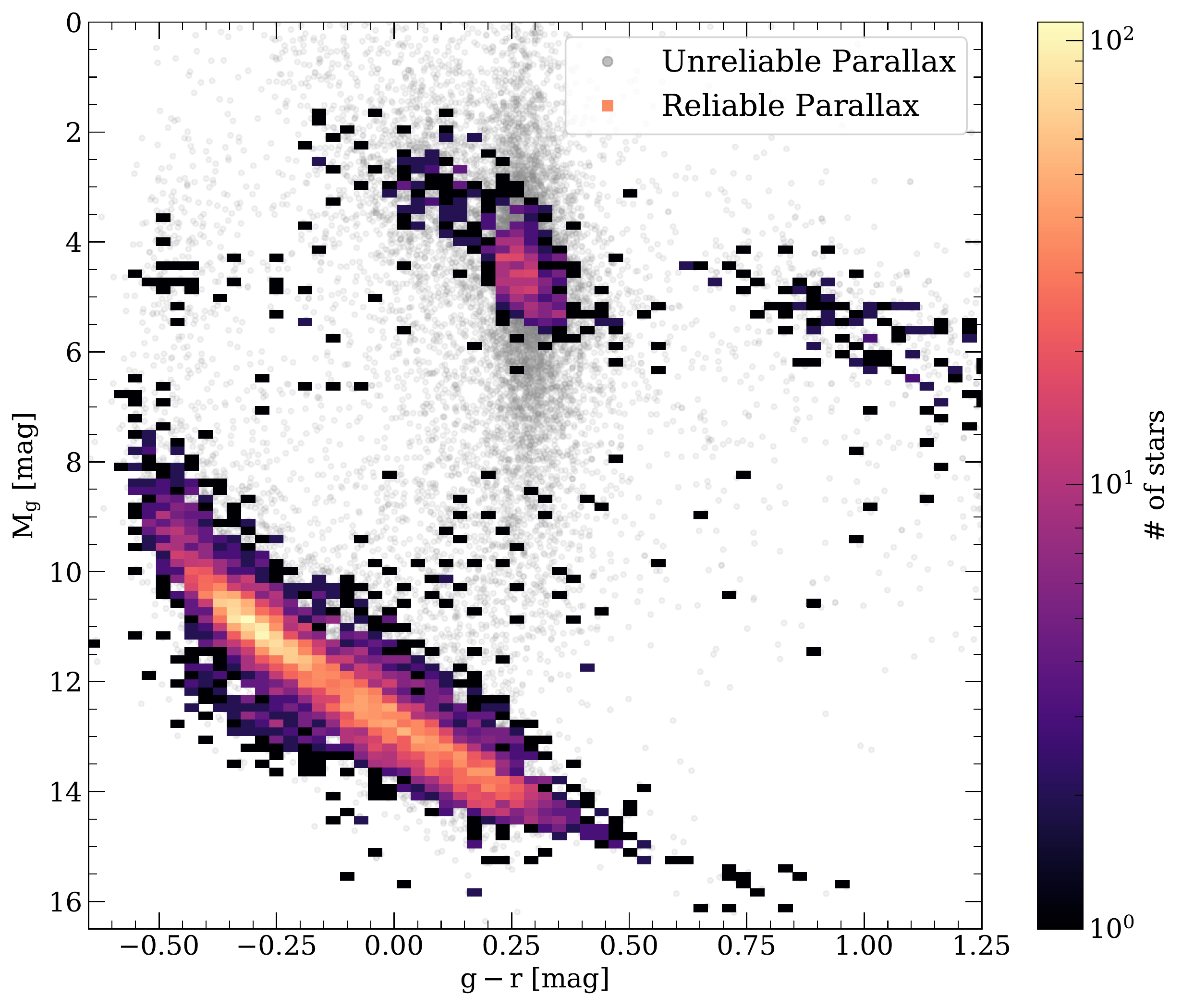}
    \caption{Color--magnitude diagram of the stars classified by
    \citet{Kepler2019a} as white dwarfs or subdwarf A stars based on
    SDSS spectroscopy.  Stars with unreliable parallaxes---75\% of the
    sample---are not included in the 2D histogram and are plotted in
    gray. There is a clear separation between the white dwarf track on the
    bottom left and the more luminous stellar main sequence at the top.
    While it is trivial to discriminate between white dwarfs and MPMS
    stars when parallaxes are available, most of the \citet{Kepler2019a}
    sample lacks reliable parallaxes in Gaia DR2.\label{fig:cmd}}
\end{figure}

As we argued above, visually separating MPMS stars from white dwarf
is prone to error, especially at cool temperatures where MPMS star
and white dwarf spectra are visually similar.  For that reason,
\cite{Kepler2019a} appealed to Gaia DR2 parallaxes to distinguish MPMS
stars and DA white dwarfs.  At a given color, white dwarfs are far less
luminous than MPMS stars due to their smaller radii.  A high-quality
parallax measurement therefore provides accurate classifications via
absolute magnitudes.  However, only about 15\% of the 14,522 stars in
this sample have reliable parallaxes (i.e., $\pi/\sigma_\pi > 10$).
As a result, the MPMS star and white dwarf classification cannot be
confirmed for most of this sample.  Consequently, when validating our
methods we only use those stars for which a definite classification
can be made based on a high-quality Gaia DR2 parallax.  We require (1)
\texttt{parallax\_over\_error > 10}, (2) \texttt{visibility\_periods\_used
> 8}, and (3) \texttt{astrometric\_sigma5d\_max < 1}.  This results in an
empirical validation sample of 1807 stars with reliable MPMS star/white
dwarf classifications.   Of these, 65\% are classified as DA white dwarfs
and the remaining 35\% are classified as MPMS stars.  We illustrate
this empirical validation sample in the Sloan color--magnitude diagram
depicted in Figure \ref{fig:cmd}.

\section{Methods}\label{sec:methods}

Even though visual classifications have been sufficient to separate MPMS
stars from white dwarfs in the past, the volume of data expected to be
produced by ongoing and next-generation spectroscopic surveys will make
the traditional approach impractical.  We therefore use the grids of
theoretical spectra described in Section \ref{sec:theorydata} to build
an automated process to differentiate MPMS stars from white dwarfs using
photometric and spectroscopic observables alone without appealing to
accurate parallax measurements.  We then validate our classifier using
the empirical labels in Section \ref{sec:empiricaldata}.

\subsection{Balmer Lines}\label{sec:theory_balmer}

Both relatively warm MPMS stars and white dwarfs exhibit strong Balmer
lines in their spectra.  On this basis, in the absence of high-quality
parallaxes MPMS stars have usually been differentiated from white
dwarfs based on Balmer lines for temperatures $T_{\text{eff}} \gtrsim
6000\text{~K}$.  As we described above, MPMS stars have narrower Balmer
lines than white dwarfs at a given temperature \citep{Kepler2019a}.
We now quantify this difference across several Balmer lines at once and
use it to produce a rigorous selection function.

For each spectrum in the theoretical grids described in Section
\ref{sec:theorydata} we fit each of the Balmer absorption lines H$\alpha$,
H$\beta$, H$\gamma$, and H$\delta$ with a Voigt profile\footnote{A
Voigt profile is a convolution of Gaussian and Lorentzian profiles
that well-approximates the pressure-broadened wings of the Balmer
lines \citep[e.g.,][]{Tremblay2009}.}.  From each profile we derive two
summary statistics: the full-width at half-maximum (FWHM) in angstroms
and the minimum of the profile in continuum-normalized flux units that
we define as the line amplitude.  For each spectrum we therefore derive
eight line statistics in total which together quantify the phase space of
the Balmer lines.  This results in a vector of eight Balmer features per
spectrum plus a ``label'' identifying it as the spectrum of a MPMS star
or white dwarf.  Our classifier then makes use of all eight features to
differentiate MPMS stars from white dwarfs. We focus on H$\alpha$,
H$\beta$, H$\gamma$, and H$\delta$ in this study because Balmer lines
bluer than H$\delta$ occur in lower signal-to-noise ratio regions of
the SDSS spectra of cool stars.

\begin{figure}
    \centering
    \includegraphics[width=\columnwidth]{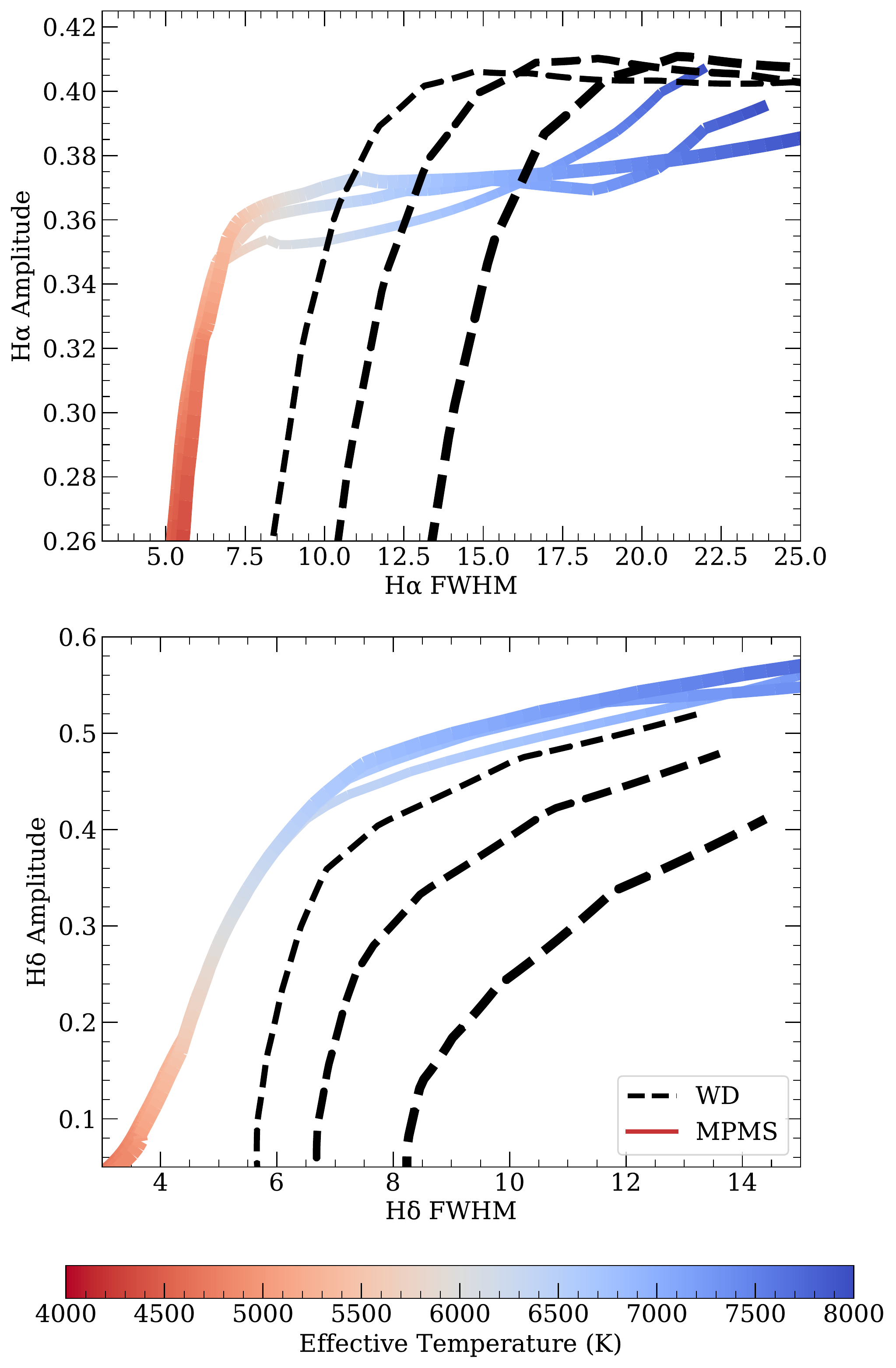}
    \caption{Two ``slices'' of the Balmer line phase space for MPMS
    stars (solid lines colored by effective temperature) and white dwarfs
    (dashed lines).  The thickness of each line is proportional to surface
    gravity $\log{g} =\ $4, 4.5, 5 for MPMS star models and $\log{g} =\
    $7.5, 8, 8.5 for white dwarf models.\label{fig:balmer_phasespace}}
\end{figure}

We illustrate in Figure \ref{fig:balmer_phasespace} two 2D ``slices''
through the Balmer line phase space that demonstrate why Balmer line
features are useful for classification.  Because of the faintness
of cooler white dwarfs, most white dwarfs targeted by the SDSS have
$T_{\text{eff}} \gtrsim 6000\text{~K}$.  As a consequence of the low
space density of massive stars in high Galactic latitude fields, most
MPMS stars targeted by the SDSS have temperatures $T_{\text{eff}}
\lesssim 8000\text{~K}$.  In the overlap region between these two
temperature ranges, the theoretical spectra confirm the expectation that
MPMS stars have much narrower (i.e., smaller FWHM) H$\alpha$ profiles
than white dwarfs.

Figure \ref{fig:balmer_phasespace} is representative of the relationship
between Balmer line features as a function of $T_{\text{eff}}$ and
$\log{g}$.  Above $T_{\text{eff}} \approx 6000\text{~K}$, MPMS star and
white dwarf spectra are visually differentiable on many 2D slices through
Balmer line phase space.  Below $T_{\text{eff}} \approx 6000\text{~K}$,
the higher-order Balmer lines in white dwarf spectra start to rapidly lose
their intensities.  In the range $5000\text{~K} \lesssim T_{\text{eff}}
\lesssim 6000\text{~K}$, combining statistics from several Balmer lines
can still separate MPMS stars from white dwarfs.  Below $T_{\text{eff}}
\approx 5000\text{~K}$, the Balmer lines for both metal-free low-mass
stars and white dwarfs begin to disappear altogether, resulting in
pure-continuum spectra \citep{Blouin2019}.  In other words, below
$T_{\text{eff}} \approx 5000\text{~K}$ there is no way to use optical
spectra to distinguish low-mass Population III stars from white dwarfs.
We discuss this special case of very low-mass Population III stars further
in Section \ref{sec:results}.
\newpage

\subsection{Synthetic Photometry}

In addition to the Balmer features we extract from each theoretical
spectrum, we also calculate synthetic photometry in several photometric
systems using the \texttt{pyphot} utility \citep{pyphot}.  We calculate
synthetic absolute magnitudes in the SDSS \textit{ugriz}, Pan-STARRS
\textit{grizy} \citep{Chambers2016}, SkyMapper \textit{uvgriz}
\citep{Bessell2011}, DECam \textit{ugrizY} \citep{Flaugher2015},
and Vera Rubin Observatory \textit{ugrizy} \citep{LSST2019} systems.
While the software accompanying this paper can be used with any of these
photometric systems, we focus on SDSS photometry colors in this paper
since we validate our method on SDSS data.

\begin{figure}
    \centering
    \includegraphics[width=\columnwidth]{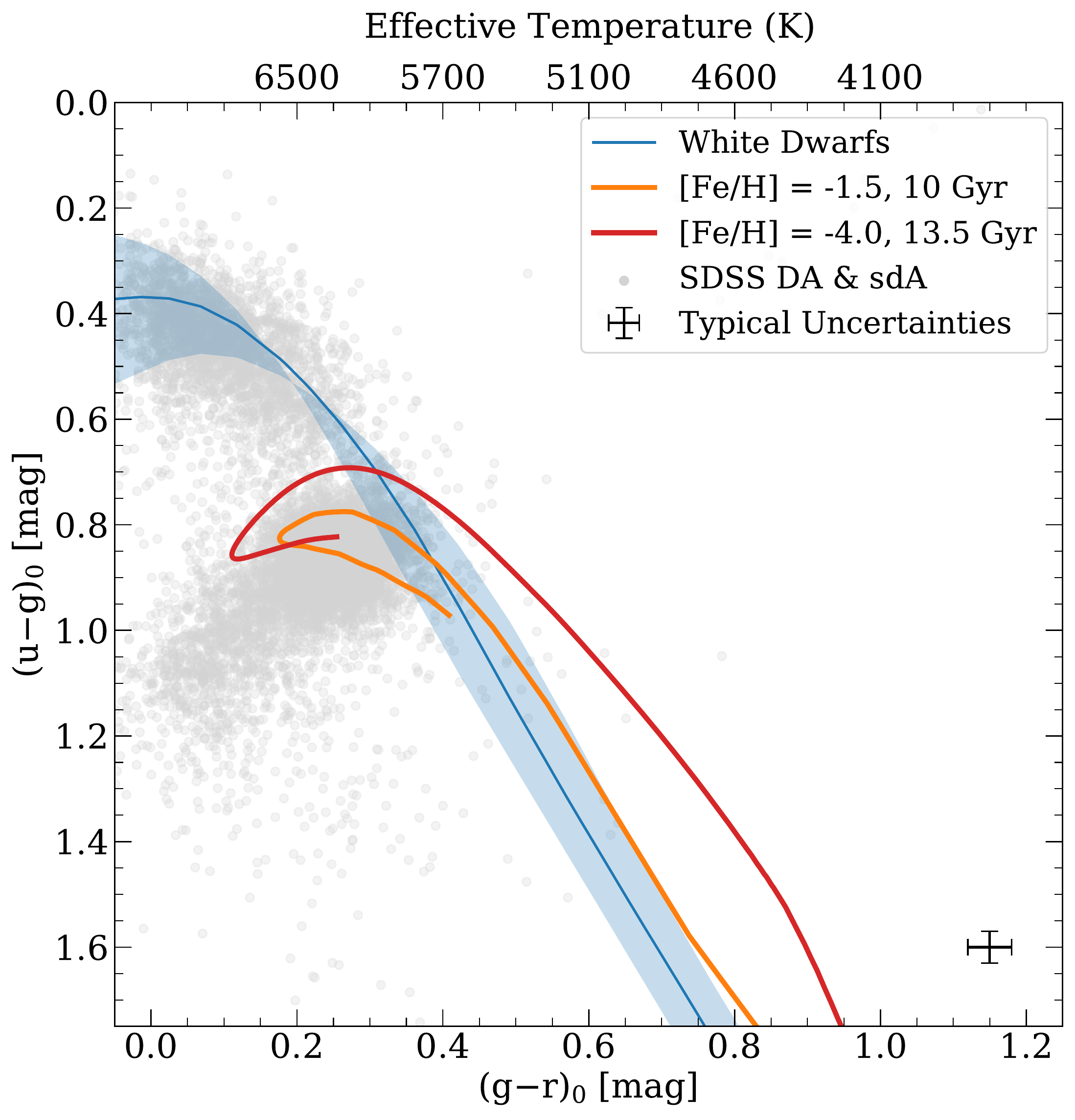}
    \caption{Synthetic color--color diagram using SDSS bands.  The blue
    line corresponds to white dwarfs with $\log{g} = 8$ while the shaded
    region indicates the locus of white dwarfs with surface gravities
    in the range $7 < \log{g} < 9$.  We also plot MIST isochrones for
    MPMS stars with $[\text{Fe/H}] = -1.5$ and age 10 Gyr (orange)
    and $[\text{Fe/H}] = -4.0$ and age 13.5 Gyr (red).  We overlay as
    gray points all stars classified as ``DA'', ``sdA'', or ``sdA/F''
    objects by \cite{Kepler2019a}.\label{fig:synth_colorcolor}}
\end{figure}

We plot in Figure \ref{fig:synth_colorcolor} a comparison between
MPMS stars and white dwarfs in Sloan color--color space.  Sloan $u-g$
color is strongly affected by the Balmer jump, a sensitive probe
of surface gravity.  A $u-g$ versus $g-r$ color--color plot can
therefore be used to cleanly differentiate relatively warm MPMS stars
and white dwarfs.  At lower temperatures the difference becomes less
pronounced, so this color--color diagram can only differentiate
MPMS stars and white dwarfs when $T_{\text{eff}} \gtrsim 6000$.
While Figure \ref{fig:synth_colorcolor} depicts one well known
photometric difference between MPMS stars and white dwarfs, we
consider the complete color--color spaces defined by each of the SDSS,
Pan-STARRS, SkyMapper, DECam, and Rubin Observatory photometric systems,
computing all unique pairwise differences of the magnitudes in each
photometric band. While the approach that we describe in the
following subsection is related to traditional color--color selections
\citep[e.g.,][]{Lokhorst2016,Pelisoli2018}, our approach uses the entire
$N$-dimensional color space distribution at once rather than relying on
individual 2D slices through that space.

\subsection{Logistic Regression Classifier}\label{sec:lr}

To map the spectroscopic and photometric features described above to a
MPMS star or white dwarf classification, we use a logistic regression
classifier.  A logistic regression is a model that assumes a linear
relationship between the input features and the log-odds of a binary
(i.e., Bernoulli) random variable taking on the value 1 (or ``true'').
Given input features $\mathbf{x} = \left[x_0, x_1, x_2,\dots\right]$,
fitting a logistic regression involves solving for the coefficients
$\bm{\beta} = \left[\beta_0, \beta_1, \beta_2,\ldots\right]$ such that
\begin{equation}
    \bm{\beta}^\intercal \mathbf{x} = \log{\left(\frac{p}{1-p}\right)},
\end{equation}
where $p$ is the probability that the Bernoulli random variable is 1
(or ``true'').  A logistic regression model is typically fit to data
with known labels (i.e., where $p$ is known to be either zero or one),
and the coefficients $\bm{\beta}$ are subsequently used to estimate
$\hat{p}$ for new input data.  There is no closed-form solution to
determine $\bm{\beta}$, so an iterative gradient descent algorithm is
often used to find the optimal coefficients.

We define the underlying Bernoulli variable to have value 0 if an object is a confirmed white dwarf and 1 if an object is
a confirmed metal-poor main-sequence star.  The associated probability
in the model can therefore be defined as $p = P_{\text{MPMS}}$,
the probability that a given object is a MPMS star as opposed to a
white dwarf.  We demonstrate three possible input configurations:
the Balmer line summary statistics (eight features), \textit{ugriz}
photometric colors (ten features), and \textit{griz} photometric colors
(six features).  We also consider a combined classifier that uses
\textit{ugriz} colors and Balmer features simultaneously.

We use a logistic regression model due to its simplicity and ease of
interpretation.  The classification probability $P_{\text{MPMS}}$ returned
by a logistic regression is well-calibrated by default \citep{Yu2011},
providing confidence in the classification.  We found that using a more
complex classification algorithm like a random forest or support vector
machine increased the complexity of the method without much increase
in accuracy.  In addition, these other algorithms require an external
``calibration function'' to transform the returned classification
probabilities to statistically meaningful values \citep{Niculescu2005}.
This step is unnecessary for the logistic regression.

When training our logistic regression models, we reserve 5\% of the
stars in our grid of theoretical models as a synthetic validation set.
We evaluate our logistic regression model on these unseen synthetic
validation data and confirm that the model correctly predicts their
classifications virtually 100\% of the time.  This validation step
ensures that the logistic regression is working as expected on
noiseless synthetic data.  We will further validate our logistic
regression model with SDSS photometry and spectroscopy for objects
with secure parallax-based classifications in Section \ref{sec:valid}.
Our classifier and synthetic training data for the SDSS, Pan-STARRS,
SkyMapper, DECam, and Rubin Observatory photometric systems are publicly
available.\footnote{\url{https://github.com/vedantchandra/mpms}\label{mpms_git}}

\subsection{Model Validation with Objects that have Secure Empirical
Classifications}\label{sec:valid}

We now further validate our classifier with the empirical validation
sample of 1807 objects described in Section \ref{sec:empiricaldata}.
These objects were classified by \citet{Kepler2019a} based on
spectroscopic features with Gaia DR2-confirmed classifications as either
MPMS stars or white dwarfs.  We confirm that the \cite{Kepler2019a}
classifications for these objects are accurate based on our inspection
of their locations in a color--magnitude diagram.  We use the available
SDSS photometry and spectroscopy for this empirical validation sample
to compute Balmer features and \textit{ugriz} colors.  We then run the
logistic regression classifier described in Section \ref{sec:lr} on these
features to predict based solely on SDSS photometry and spectroscopy
the probability that each object is a MPMS star.

Receiver operating characteristic (ROC) curves are one way to graphically
evaluate classifiers.  ROC curves describe the relationship between the
true-positive and false-positive rates of a classifier as a function
of its acceptance threshold (i.e., the output probability above
which a ``positive'' classification is made).  ROC curves provide a
high-level summary of the sensitivity (i.e., probability of detection)
and specificity (i.e., probability of false alarm).  Precision-recall
(PR) curves provide an alternative graphical classifier diagnostic.
They are particularly useful for our application, as precision (i.e., the
probability that a MPMS star is accurately classified by our classifier)
is the most important metric for our scientific goal.  PR curves are
also more agnostic to the possible existence of class imbalance in a
data set, as is the case in our empirical validation sample in which
there are twice as many white dwarfs as MPMS stars.  Together, ROC and
PR curves provide an overview of the performance of a classifier.

\begin{figure}
    \centering
    \includegraphics[width=\columnwidth]{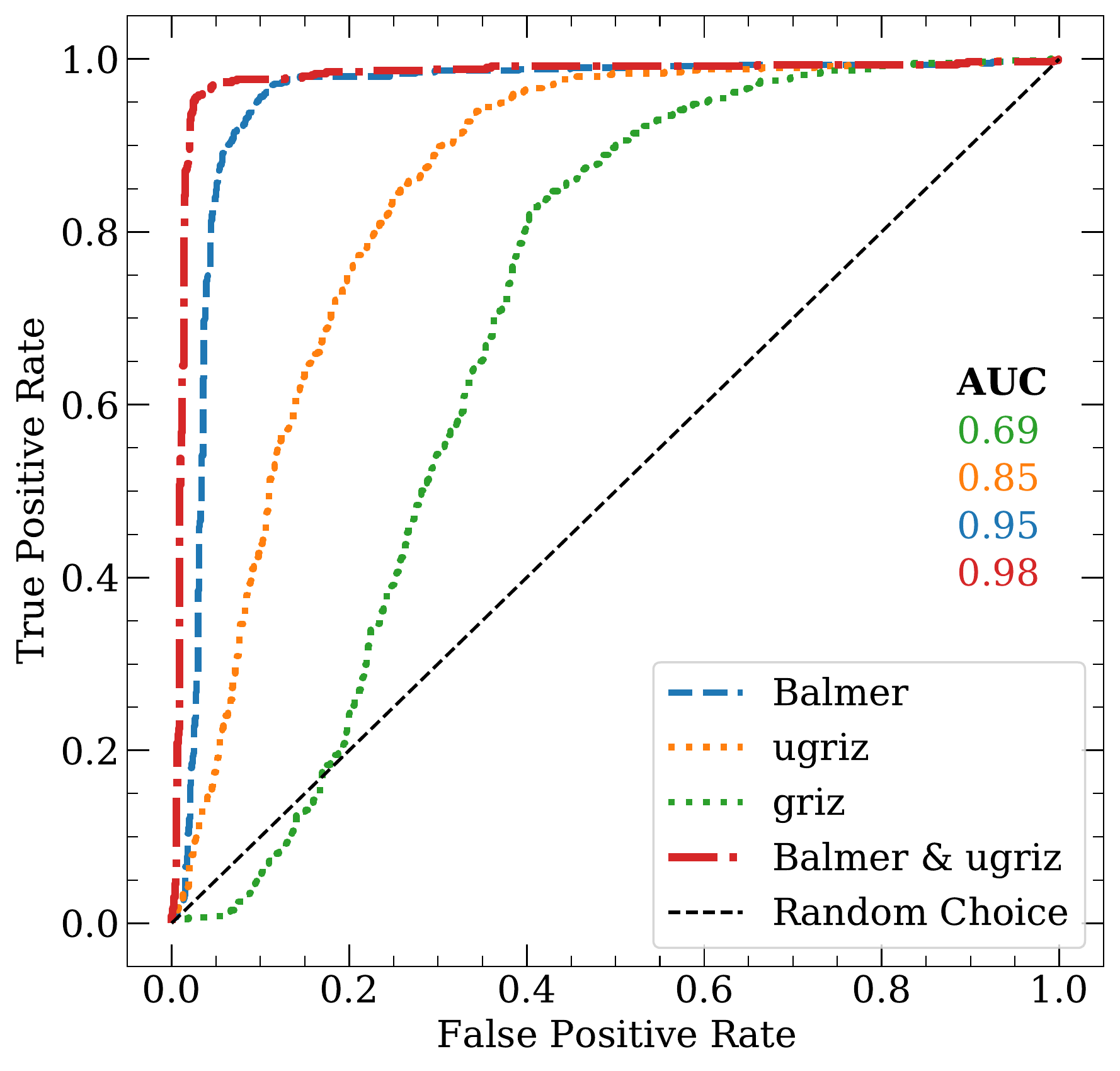}
    \includegraphics[width=\columnwidth]{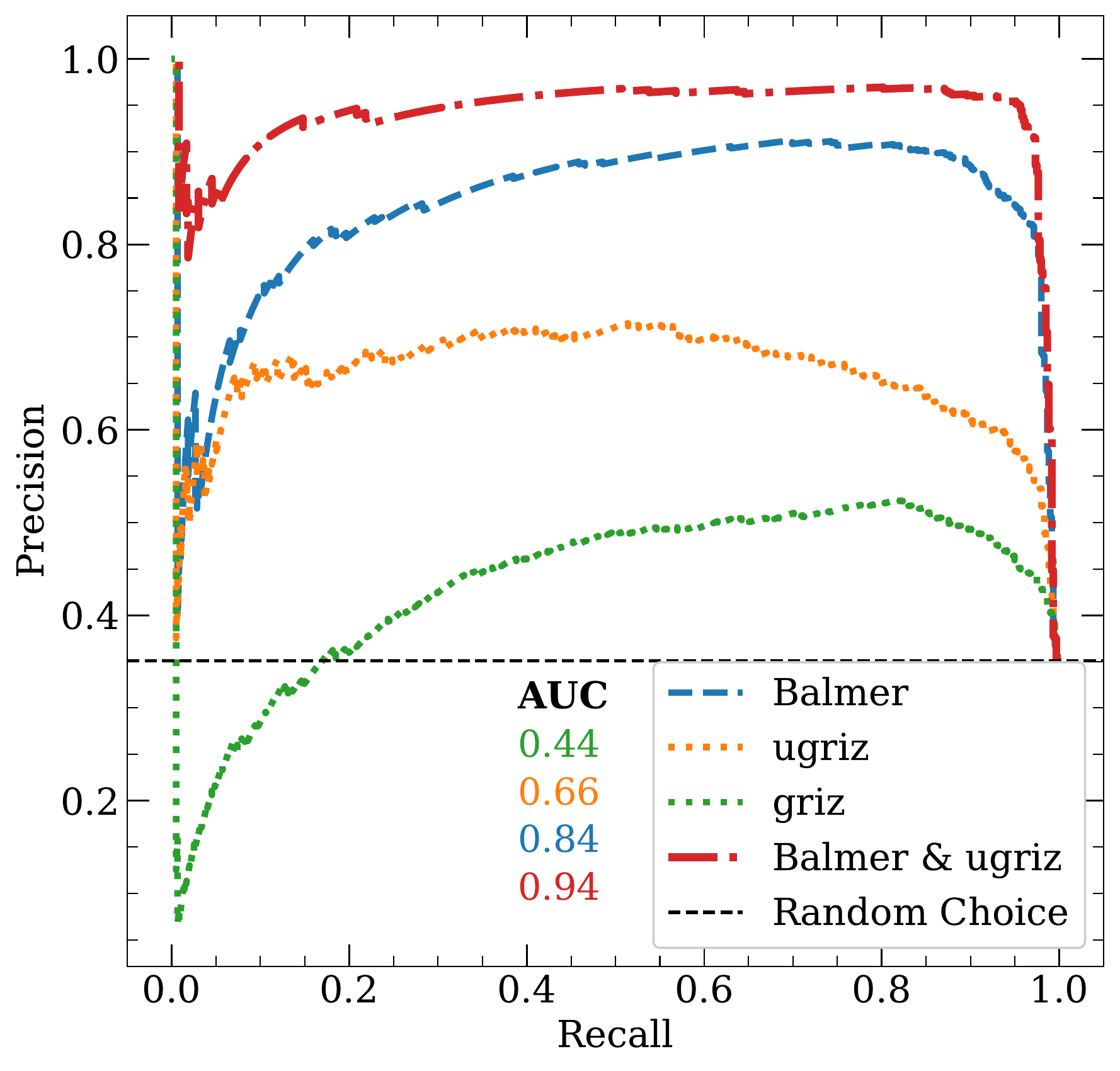}
    \caption{Receiver operating characteristic (ROC) and precision-recall
    (PR) curves for the photometry-only, spectroscopy-only, and
    combined logistic regression classifiers trained on theoretical
    models and validated with classifications confirmed by high-quality
    Gaia DR2 parallaxes.  The classifier that uses both spectroscopic
    and photometric features achieves the best performance, followed
    closely by the classifier that uses only spectroscopic features.
    Classifiers that use only photometric features achieve acceptable
    performance but lag behind classifiers that utilize spectroscopic
    features.\label{fig:roc_validation}}
\end{figure}

We illustrate in Figure \ref{fig:roc_validation} ROC and PR curves for
the application of our classifiers to our empirical validation data.
We evaluate our logistic regression classifier in several scenarios:
only the eight Balmer-line-summary-based features, only the ten
\textit{ugriz} colors, only the six \textit{griz} colors, and both
the Balmer-line-summary-based features and \textit{ugriz} colors.
As expected, the Balmer spectroscopic features provide the most
discriminating power.  The inclusion of $u-g$ color greatly improves
the quality of classifications compared to $griz$ colors alone, as $u-g$
color is sensitive to $\log{g}$ via the Balmer break.

In terms of precision, the classifier using both photometric and
spectroscopic features far outperforms the spectroscopy-only classifier.
This implies that a degeneracy in spectroscopic features in MPMS star
and white dwarf spectra can be overcome by including photometric
features. That natural interpretation is that the photometric
colors provide a strong constraint on temperature, breaking the
$T_{\text{eff}}$--$\log{g}$ degeneracy of the absorption line
features. We validate the naive expectation that classifiers
using more than one Balmer line are superior to any classifier based
on a single line.  We find that classification performance is most
enhanced when lower-order Balmer lines like H$\alpha$ are combined with
higher-order lines like H$\delta$ and confirm that the best performance
is achieved when all four Balmer lines are included.  We conclude that
our combined photometric and spectroscopic classifier has the accuracy
and precision necessary to separate MPMS stars from white dwarfs using
SDSS data or equivalent LAMOST or DESI data.

While our photometry-only classifier can be used to separate
MPMS stars from white dwarfs, it is susceptible to many of the issues
with traditional color--color selections.  For example, the colors of
white dwarfs with heavily metal-polluted photospheres will change due to
line blanketing and changes in continuum opacity.  While metal-polluted
white dwarfs could be misclassified by our photometry-only classifier,
they are easy to remove from a joint photometric/spectroscopic search
for metal-poor stars due to their prominent metal absorption lines.
Our classifier is therefore most powerful when spectroscopic data are
available.

\subsection{Approximate Determination of Metallicity}\label{sec:ew}

After differentiating MPMS stars from white dwarfs, another important
task is to select the most metal-poor candidates among the predicted
MPMS stars.  In order to select the most metal-poor MPMS star candidates,
we implement a simple method to approximately quantify stellar metallicity
from MPMS star spectra by fitting them with theoretical spectra.

We first use SDSS photometry and any available Gaia
parallax information to generate an initial set of spectroscopic
stellar parameters $T_{\text{eff}}$, $\log{g}$, and $[\text{Fe/H}]$.
We define a $\chi^2$ statistic by comparing the observed SDSS $ugriz$
photometry dereddened as described in Section \ref{sec:empiricaldata}
to MESA Isochrones \& Stellar Tracks (MIST) $ugriz$ photometry
\citep{Dotter2016,Choi2016,Paxton2011,Paxton2013,Paxton2015,Paxton2018}.
To account for the low-precision parallaxes available in most
cases, we use an exponentially decreasing space density prior on
parallax \citep{Bailer-Jones2015}. We sample with \texttt{emcee}
\citep{Foreman-Mackey2013,Foreman-Mackey2019} the posterior distribution
of the fundamental stellar parameters (e.g., mass, age, and metallicity)
conditioned on the data, taking $-0.5 \cdot \chi^2$ as the log-likelihood of the data.  We then identify the maximum-likelihood
sample and interpolate the MIST isochrone grid to define an initial
set of spectroscopic stellar parameters $T_{\text{eff}}$, $\log{g}$,
and $[\text{Fe/H}]$.

Next, we fit PHOENIX theoretical spectra from \citet{Husser2013}
spanning the range $3000~\text{K} \leq T_{\text{eff}} \leq 8000~\text{K}$,
$3.5 \leq \log{g} \leq 6.5$, and $-4 \leq [\text{Fe/H}] \leq 0$ to the
observed SDSS MPMS star spectra using a a two-step fitting procedure. We initialize the stellar parameters at the photometrically-inferred values described above.
We first fit the sky emission line at 4047 \AA\ in each SDSS spectrum with
a Gaussian to estimate the instrumental line-spread function.  We then
convolve the grid of theoretical spectra with this kernel to bring them
to the instrumental resolution.  We next fit a straight line to a region
around H$\alpha$ in each SDSS spectrum and then divide the flux by that
line to approximately normalize the spectrum in the vicinity of the line.
We then fit a Gaussian to H$\alpha$ and use its known wavelength to
correct each SDSS spectrum to the rest frame.  To spectroscopically infer
$T_{\text{eff}}$ and $\log{g}$, we assume a fixed $[\text{Fe/H}] = -2$
and fit the grid of theoretical spectra to the continuum-normalized
Balmer lines.  To spectroscopically infer $[\text{Fe/H}]$, we fix
the $T_{\text{eff}}$ and $\log{g}$ derived in the previous step and
fit the continuum-normalized \ion{Ca}{2} K absorption line to infer
$[\text{Ca/H}]$ and thereby $[\text{Fe/H}]$ assuming $[\text{Ca/Fe}]
= 0$.  For each absorption line, we fit the theoretical spectra using a
non-linear least squares algorithm \texttt{lmfit} \citep{Newville2014}
that minimizes the squared residual between the observed and theoretical
spectra.  We alternate the Balmer-based $T_{\text{eff}}$ \& $\log{g}$
and the \ion{Ca}{2} K-based $[\text{Fe/H}]$ inferences until the
spectroscopic stellar parameters are self-consistent. In practice this only requires one or two iterations since the Balmer-based inference is quite independent of metallicity.  

For the metal-poor
main sequence stars with published spectroscopic stellar parameters from
higher resolution or signal-to-noise ratio (S/N) data, we find that the
spectroscopically derived parameters are better than those based on photometry and astrometry. This is likely due to the
low-quality parallaxes available in most cases. Additionally, we note that spectroscopic $\text{Fe/H}$ inferences based on the \ion{Ca}{2} K absorption line can be overestimated if the star is carbon-enhanced, since the \ion{Ca}{2} K line is blended with nearby carbon features on the mid-resolution spectra \citep[e.g.,][]{Frebel2015a}. Since our primary goal is to find the most metal-poor candidates (including those depleted in carbon), we do not correct for this effect explicitly, but we caution that our spectroscopically-inferred metallicities should be treated as indicative estimates only.

\section{Results}\label{sec:results}

With the accuracy and precision of our classifier now established, we
can now classify the entire sample of 14,522 MPMS star/white dwarf candidates
from \cite{Kepler2019a} with SDSS photometry and spectroscopy.  We fit the Balmer
lines of these spectra to derive the line summaries described in
Section \ref{sec:theory_balmer}.  We then apply our combined logistic
regression classifier to the Balmer features and $ugriz$ colors of these
stars.  The result of this process is a prediction of the probability
$P_{\text{MPMS}}$ that a spectrum was produced by a MPMS star as opposed
to a white dwarf.  Reassuringly, the distribution of probabilities
is bimodal: most stars have probabilities close to zero or one.
This distribution provides even more confidence that our classifier is
both precise and accurate.  We decide on decision threshold of 0.5 and
classify all objects with $P_{\text{MPMS}} > 0.5$ as MPMS star candidates.

We find that approximately 95\% of our logistic regression-based
classifications agree with the visual inspection-based classifications
reported in \citep{Kepler2019a}.  The 5\% of cases where our classifier
disagrees with \cite{Kepler2019a} are due to objects with extremely low
S/N spectra.  Spectroscopic surveys like DESI will produce hundreds of
thousands of spectra plausibly produced by MPMS stars or white dwarfs,
and only our automated classification approach will be able to scale to
that data volume.

We emulate future searches for candidate extremely metal-poor
(EMP) stars in upcoming surveys like SDSS-V and DESI by mining
the stars in our SDSS data set for EMP candidates.  Numerous
studies have already searched these SDSS data for metal-poor stars
\citep[e.g.,][]{Allende_Prieto2006,Allende_Prieto2008,Allende_Prieto2014,Allende_Prieto2015,Caffau2011,Caffau2011b,Bonifacio2012,Bonifacio2015,Aoki2013,Frebel2015b,Placco2015,Aguado2016,Aguado2017a,Aguado2017b,Aguado2018b,Aguado2018a,Yoon2016,Carbon2017},
so our expectation is that many of the most promising candidates
have already been discovered.  We select all stars from the 12,000
star subsample of the \citet{Kepler2019a} study referenced above
with $P_{\text{MPMS}} > 0.5$ according to our combined classifier as
possible MPMS candidates.  This leaves us with 10,500 MPMS star spectra for
which we compute approximate metallicities
as described in Section \ref{sec:ew}. We selected stars with the lowest
estimated metallicities and visually inspected their spectra, examining
the \ion{Ca}{2} K (3934.8 \AA), \ion{Na}{1} (5895.6 \AA), and \ion{Mg}{1}
(5176.7 \AA) lines in particular.  We report our most promising candidates
in Table \ref{tab:emp_candidates}, along with our spectroscopically-inferred estimates of $T_{\text{eff}}$, $\log{g}$, and $\text{[Fe/H]}$.

\begin{deluxetable*}{ccccccccc}
\centerwidetable
\tablewidth{\textwidth}
\tablecaption{Extremely Metal-poor Candidates Identified in This Study}
\startdata
\tablehead{
\colhead{Gaia DR2} & \colhead{R.A.} & \colhead{Decl.} & \colhead{Sloan} & \colhead{$g$} & \colhead{$T_{\text{eff}}$} & \colhead{$\log{g}$} & \colhead{[Fe/H]}\vspace{-5pt} & \colhead{Reference} \\
\colhead{Source ID} & \colhead{(deg)} & \colhead{(deg)} & \colhead{Plate-MJD-Fiber} & \colhead{(mag)} & \colhead{(K)} & \colhead{} & \colhead{}}
3724067121891339648 & 211.826 & 10.505 & 1703-53799-0013 & 16.9 & 6441 & 4.0 & -4.0 &  \\
874891332185786752 & 116.953 & 26.762 & 2055-53729-0121 & 17.3 & 6450 & 3.7 & -4.0 &  \\
4465215302785008640 & 245.076 & 16.141 & 2188-54595-0026 & 18.5 & 6550 & 4.1 & -3.9 &  \\
1448863790892447744 & 203.103 & 27.516 & 2245-54208-0450 & 16.8 & 6435 & 4.0 & -3.8 &  \\
3958825597588943232 & 189.109 & 23.019 & 3374-54948-0220 & 18.2 & 6453 & 3.7 & -3.8 &  \\
3905214200892445184 & 184.325 & 8.870 & 5396-55947-0370 & 19.4 & 6560 & 4.5 & -3.8 &  \\
1493736818927192576 & 220.706 & 44.954 & 6046-56096-0550 & 18.8 & 6359 & 4.0 & -3.8 &  \\
\tableline
3092216989574463616 & 126.339 & 4.059 & 1185-52642-0519 & 17.1 & 6441 & 4.0 & -3.9 & \cite{Aoki2013} \\
290930261314166528 & 25.151 & 23.749 & 2044-53327-0515 & 15.8 & 6354 & 4.8 & -3.5 & \cite{Arentsen2019} \\
1276882477044162688 & 230.509 & 30.924 & 1651-53442-0330 & 16.6 & 6333 & 4.7 & -3.4 & \cite{Aoki2013} \\
4190837398756490112 & 301.306 & -10.751 & 2303-54629-0377 & 17.0 & 6536 & 3.8 & -3.6 & \cite{Francois2018} \\
1184737183522291712 & 221.669 & 12.822 & 1712-53531-0636 & 16.2 & 6445 & 4.0 & -2.9 & \cite{Aoki2013} \\
3740963179636227968 & 207.345 & 14.127 & 1777-53857-0479 & 16.6 & 6340 & 3.8 & -3.5 & \cite{Bonifacio2018}  \\
3976087728282022272 & 175.848 & 20.349 & 2506-54179-0576 & 16.9 & 6426 & 4.0 & -2.9 & \cite{Yoon2016} \\
3890626773968983296 & 157.313 & 17.491 & 2853-54440-0113 & 16.9 & 6047 & 4.7 & -4.0 & \cite{Caffau2011} \\
1609914447333432192 & 213.030 & 56.159 & 2447-54498-0274 & 16.0 & 6488 & 3.8 & -4.0 & \cite{Aoki2013} \\
1195572458297534080 & 233.443 & 15.950 & 2782-54592-0411 & 16.9 & 6562 & 4.0 & -4.0 & \cite{Caffau2013} \\
931227322991970560 & 123.976 & 47.496 & 3693-55208-0408 & 17.1 & 6523 & 4.6 & -4.0 & \cite{Aguado2018a} \\
2548541852945056896 & 5.808 & 3.133 & 4299-55827-0002 & 17.9 & 6404 & 4.0 & -4.0 & \cite{Aguado2018b} \\
\enddata
\tablecomments{We indicate SIMBAD references for stars with confirmed $\text{[Fe/H]} < -3.0$ inferences. $T_{\text{eff}}$, $\log{g}$ and $\text{[Fe/H]}$ are estimated from Sloan spectra with the spectroscopic fitting routine described in Section \ref{sec:ew}. \label{tab:emp_candidates}
}
\end{deluxetable*}

In addition, we use SIMBAD to search for literature references to our
candidate MPMS stars \citep{Wenger2000}.  While several of our MPMS
star candidates have no prior reference in the literature, we find that
our selection turns up dozens of previously-known metal-poor stars
with confirmed metallicities $[\text{Fe/H}] < -3.0$. A selection of these stars is included in Table \ref{tab:emp_candidates}. We therefore
conclude that SDSS data have already been exhaustively searched for
extremely metal-poor stars.  Nevertheless, this serves as a comprehensive
validation of our methodology to search for metal-poor stars in future
spectroscopic surveys.

While the existence of low-mass Population III stars is still debated, we
assert that it will be possible to uniquely classify a featureless optical
spectrum plausibly produced be either a low-mass Population III star or a
cool white dwarf based on the optical-to-infrared colors of the object.
The lowest mass primordial-composition object that can burn hydrogen
should have $M_{\ast} \approx 0.1~M_{\odot}$.  Stellar models therefore
predict a lower limit on the effective temperature of Population III stars
of $T_{\text{eff}} \approx 3600\text{~K}$ \citep{Burrows1993,Saumon1994}.
At these temperatures, the optical spectra of the coolest primordial stars
will be indistinguishable from the optical spectra of white dwarfs because
both will have pure-continuum spectra with no Balmer or metal lines.

However, collision-induced absorption (CIA) from molecular hydrogen
will produce surface gravity-dependent continuum features in the red
optical and infrared \citep{Saumon1994,Blouin2017,Blouin2019}.  At a
given temperature (or $g-r$ color), a white dwarf will have more CIA
than a low-mass Population III star due to its higher surface gravity.
White dwarfs will consequently be fainter than low-mass Population III
stars in the red optical $izy$ and infrared bands like $JHK$.  Therefore,
at constant temperature photometric colors comparing optical to red
optical or infrared bands (e.g., $r-z$, $V-J$, etc.) will be bluer
for white dwarfs than low-mass Population III stars.  This effect
is expected to be larger than the typical photometric uncertainties
in ground-based surveys like the SDSS, Pan-STARRS, SkyMapper, or DES
\citep{Saumon1994,Blouin2017}.

While future work on CIA opacities will enable a more quantitative
comparison, current uncertainties in the theoretical modeling of CIA
make it challenging to precisely quantify the expected difference
between low-mass Population III stars and white dwarfs, and thereby
build a predictive model like the one we presented for warmer stars.
Nevertheless, we predict that the photometric signature of CIA should be
detectable via infrared photometry and consequently able to differentiate
between low-mass Population III stars and white dwarfs even without
parallax information.

\section{Discussion}\label{sec:discussion}

We have developed a classification framework that can be used
to differentiate low-mass Population III stars from white dwarfs in
future spectroscopic surveys.  Since our techniques primarily rely on
Balmer lines and broadband photometry, our methods naturally extend
to the identification of metal-poor stars including the sought-after
EMP and ultra metal-poor (UMP) stars (i.e., stars with $[\text{Fe/H}]
\lesssim -4$).  Another natural application of our technique is the
identification and removal of the subdwarf stars that frequently
contaminate catalogs of white dwarf candidates \citep{Kepler2019a}.

The typical white dwarf has $M_{\ast} \approx 0.6~M_{\odot}$, $R_{\ast}
\approx R_{\oplus}$, and consequently $\log{g} \sim 8$.  Its surface
gravity is therefore orders of magnitude larger than the typical
metal-poor main sequence star surface gravity $\log{g} \lesssim 5$.
As a result, our classifier gains most of its discriminatory power from
an object's surface gravity.  While MPMS stars are well separated from
typical white dwarfs in $\log{g}$, MPMS stars can overlap in $\log{g}$
with ELM white dwarfs produced by mass transfer in multiple systems
\citep[e.g.,][]{Brown2010,Pelisoli2016,Pelisoli2019a,Pelisoli2019b}.
Because of their lower masses and larger radii than typical white dwarfs,
ELM white dwarfs can have lower $\log{g}$ and much narrower Balmer lines
than typical white dwarfs.  Their narrow Balmer lines could cause our
classifier to misidentify them as MPMS stars.  While the identification
of these relatively rare ELM white dwarfs is a worthwhile goal in its own
right, it is not the main objective of our current analysis.  We argue
that in future searches for MPMS or low-mass Population III stars, ELM
white dwarfs can be filtered out by directly fitting theoretical spectra
to Balmer lines to infer $\log{g}$.  Indeed, even the least-massive
ELM white dwarfs should have $\log{g} \gtrsim 5$ while MPMS stars
with $M_{\ast} \gtrsim 0.1~M_{\odot}$ will have $\log{g} \lesssim 5$
\citep{Brown2010,Dotter2016,Choi2016}.

It remains an open question whether or not low-mass Population III
stars exist in the local Universe.  Beyond awaiting serendipitous direct
detections, Galactic/stellar archaeology provide a promising technique to
probe the hypothesis of surviving Population III stars and to constrain
the primordial initial mass function.  \cite{Hartwig2015a} proposed
that a total sample size of order $10^7$ halo stars is required to rule
out Population III survivors at the 95\% confidence level.  However,
this is likely an overestimate since it only considers blind surveys.
In practice, searching for EMP and UMP stars is far more efficient.
\cite{Magg2019} suggested instead to use the occurrence of EMP and
UMP stars to constrain the existence of Population III survivors.
The largest uncertainty in this method by far is the total number
of EMP and UMP stars in the Milky Way's stellar halo.  As a result,
a comprehensive search for such stars is an important need.

\section{Conclusion}\label{sec:conclusion}

While it is trivial to separate metal-poor main sequence stars and white
dwarfs when high-quality parallaxes are available, even post-Gaia many
objects with DESI or Sloan spectroscopy will lack reliable parallaxes.
For that reason, we developed a classifier capable of separating
metal-poor main sequence stars from cool white dwarfs using a range of
photometric and spectroscopic features.  We trained and validated our
classifier using theoretical spectra and synthetic photometry.  We also
validated the classifier using objects securely classified as metal-poor
main sequence stars or white dwarfs based on Sloan spectroscopy and
high-quality Gaia DR2 parallaxes.  We then applied our classifier to
a sample of candidate metal-poor main sequence stars and white dwarfs
with visual classifications and confirmed that our automated approach
reproduces the human classifications.  We make our classifier and
its underlying source code publicly available\footref{mpms_git} for reproducibility
and application to future survey data.  For the stars classified as
metal-poor main sequence stars, we executed a search for extremely
metal-poor stars.  We uncovered some previously unidentified candidates
and flagged dozens of already confirmed extremely metal-poor stars.
We predict that the application of our classifier to future DESI and
SDSS-V spectroscopy will discover thousands of metal-poor main sequence
stars.  We further predict that even in the absence of high-quality
parallaxes, any candidate low-mass Population III stars identified by
their featureless optical spectra can be separated from cool white dwarfs
due to their redder optical-to-infrared colors.  These red colors are
produced by relatively weak collision-induced absorption from molecular
hydrogen in the relatively low-surface gravity photospheres of low-mass
Population III stars. The methods presented in this work will facilitate
the rapid and reliable identification of MPMS stars, providing improved
constraints on the uncertain existence of surviving pristine stars from
the primordial Universe.

\acknowledgements
We thank the anonymous referee for discerning suggestions that
significantly improved both our analysis and this paper.  We are grateful
to Simon Blouin for providing his latest grid of theoretical cool white
dwarf spectra as well as insightful comments and suggestions.  We thank
JJ Hermes, Vinicius Placco, Pier-Emmanuel Tremblay, Tilman Hartwig,
Ingrid Pelisoli, and Davide Aguado for constructive conversations
that significantly improved this paper. This material is based upon
work supported by the Johns Hopkins University the Institute for Data
Intensive Engineering \& Science (IDIES).

This research has made use
of the SIMBAD database \citep{Wenger2000}, operated at CDS, Strasbourg,
France. This research has made use of NASA's Astrophysics Data System
(ADS). Funding for the SDSS and SDSS-II has been provided by the Alfred
P. Sloan Foundation, the Participating Institutions, the National Science
Foundation, the U.S. Department of Energy, the National Aeronautics
and Space Administration, the Japanese Monbukagakusho, the Max Planck
Society, and the Higher Education Funding Council for England. The SDSS
Web Site is \url{http://www.sdss.org/}.  The SDSS is managed by the
Astrophysical Research Consortium for the Participating Institutions. The
Participating Institutions are the American Museum of Natural History,
Astrophysical Institute Potsdam, University of Basel, University of
Cambridge, Case Western Reserve University, University of Chicago,
Drexel University, Fermilab, the Institute for Advanced Study, the Japan
Participation Group, Johns Hopkins University, the Joint Institute for
Nuclear Astrophysics, the Kavli Institute for Particle Astrophysics and
Cosmology, the Korean Scientist Group, the Chinese Academy of Sciences
(LAMOST), Los Alamos National Laboratory, the Max-Planck-Institut f\"ur
Astronomie (MPIA Heidelberg), the Max-Planck-Institut f\"ur Astrophysik
(MPA Garching), New Mexico State University, Ohio State University,
University of Pittsburgh, University of Portsmouth, Princeton University,
the United States Naval Observatory, and the University of Washington.
Funding for SDSS-III has been provided by the Alfred P. Sloan Foundation,
the Participating Institutions, the National Science Foundation, and the
U.S. Department of Energy Office of Science. The SDSS-III web site is
\url{http://www.sdss3.org/}.  SDSS-III is managed by the Astrophysical
Research Consortium for the Participating Institutions of the SDSS-III
Collaboration including the
University of Arizona, the Brazilian Participation Group, Brookhaven
National Laboratory, Carnegie Mellon University, University of Florida,
the French Participation Group, the German Participation Group, Harvard
University, the Instituto de Astrof\'isica de Canarias, the Michigan
State/Notre Dame/JINA Participation Group, Johns Hopkins University,
Lawrence Berkeley National Laboratory, Max-Planck-Institut f\"ur
Astrophysik (MPA Garching), Max-Planck-Institut f\"ur Extraterrestrische
Physik (MPE), New Mexico State University, New York University, Ohio State
University, Pennsylvania State University, University of Portsmouth,
Princeton University, the Spanish Participation Group, University
of Tokyo, University of Utah, Vanderbilt University, University of
Virginia, University of Washington, and Yale University.  Funding for
the Sloan Digital Sky Survey IV has been provided by the Alfred P. Sloan
Foundation, the U.S.  Department of Energy Office of Science, and the
Participating Institutions.  SDSS-IV acknowledges support and resources
from the Center for High Performance Computing  at the University of
Utah. The SDSS website is \url{www.sdss.org}.  SDSS-IV is managed by
the Astrophysical Research Consortium for the Participating Institutions
of the SDSS Collaboration including the Brazilian Participation Group,
the Carnegie Institution for Science, Carnegie Mellon University, Center
for Astrophysics | Harvard \& Smithsonian, the Chilean Participation
Group, the French Participation Group, Instituto de Astrof\'isica de
Canarias, Johns Hopkins University, Kavli Institute for the Physics and
Mathematics of the Universe (IPMU) / University of Tokyo, the Korean
Participation Group, Lawrence Berkeley National Laboratory, Leibniz
Institut f\"ur Astrophysik Potsdam (AIP),  Max-Planck-Institut f\"ur
Astronomie (MPIA Heidelberg), Max-Planck-Institut f\"ur Astrophysik
(MPA Garching), Max-Planck-Institut f\"ur Extraterrestrische Physik
(MPE), National Astronomical Observatories of China, New Mexico State
University, New York University, University of Notre Dame, Observat\'ario
Nacional / MCTI, The Ohio State University, Pennsylvania State University,
Shanghai Astronomical Observatory, United Kingdom Participation Group,
Universidad Nacional Aut\'onoma de M\'exico, University of Arizona,
University of Colorado Boulder, University of Oxford, University of
Portsmouth, University of Utah, University of Virginia, University of
Washington, University of Wisconsin, Vanderbilt University, and Yale
University.  This work has made use of data from the European Space
Agency (ESA) mission Gaia (\url{https://www.cosmos.esa.int/gaia}),
processed by the Gaia Data Processing and Analysis Consortium (DPAC,
\url{https://www.cosmos.esa.int/web/gaia/dpac/consortium}).  Funding for
the DPAC has been provided by national institutions, in particular the
institutions participating in the Gaia Multilateral Agreement.

\facilities{Gaia, Sloan}

\software{\texttt{numpy} \citep{Harris2020},
          \texttt{scipy} \citep{Virtanen2020},
          \texttt{matplotlib} \citep{Hunter2007},
          \texttt{astropy} \citep{Robitaille2013},
          \texttt{scikit-learn} \citep{Pedregosa2011},
          \texttt{lmfit} \citep{Newville2014},
          \texttt{mwdust} \citep{Bovy2016}}

\bibliography{library.bib}

\begin{thebibliography}{}
\expandafter\ifx\csname natexlab\endcsname\relax\def\natexlab#1{#1}\fi
\providecommand{\url}[1]{\href{#1}{#1}}
\providecommand{\dodoi}[1]{doi:~\href{http://doi.org/#1}{\nolinkurl{#1}}}
\providecommand{\doeprint}[1]{\href{http://ascl.net/#1}{\nolinkurl{http://ascl.net/#1}}}
\providecommand{\doarXiv}[1]{\href{https://arxiv.org/abs/#1}{\nolinkurl{https://arxiv.org/abs/#1}}}

\bibitem[{{Abel} {et~al.}(2000){Abel}, {Bryan}, \& {Norman}}]{Abel2000}
{Abel}, T., {Bryan}, G.~L., \& {Norman}, M.~L. 2000, \apj, 540, 39,
  \dodoi{10.1086/309295}

\bibitem[{{Abel} {et~al.}(2002){Abel}, {Bryan}, \& {Norman}}]{Abel2002}
---. 2002, Science, 295, 93, \dodoi{10.1126/science.295.5552.93}

\bibitem[{{Aguado} {et~al.}(2016){Aguado}, {Allende Prieto}, {Gonz{\'a}lez
  Hern{\'a}ndez}, {Carrera}, {Rebolo}, {Shetrone}, {Lambert}, \&
  {Fern{\'a}ndez-Alvar}}]{Aguado2016}
{Aguado}, D.~S., {Allende Prieto}, C., {Gonz{\'a}lez Hern{\'a}ndez}, J.~I.,
  {et~al.} 2016, \aap, 593, A10, \dodoi{10.1051/0004-6361/201628371}

\bibitem[{{Aguado} {et~al.}(2018{\natexlab{a}}){Aguado}, {Allende Prieto},
  {Gonz{\'a}lez Hern{\'a}ndez}, \& {Rebolo}}]{Aguado2018b}
{Aguado}, D.~S., {Allende Prieto}, C., {Gonz{\'a}lez Hern{\'a}ndez}, J.~I., \&
  {Rebolo}, R. 2018{\natexlab{a}}, \apjl, 854, L34,
  \dodoi{10.3847/2041-8213/aaadb8}

\bibitem[{{Aguado} {et~al.}(2017{\natexlab{a}}){Aguado}, {Allende Prieto},
  {Gonz{\'a}lez Hern{\'a}ndez}, {Rebolo}, \& {Caffau}}]{Aguado2017a}
{Aguado}, D.~S., {Allende Prieto}, C., {Gonz{\'a}lez Hern{\'a}ndez}, J.~I.,
  {Rebolo}, R., \& {Caffau}, E. 2017{\natexlab{a}}, \aap, 604, A9,
  \dodoi{10.1051/0004-6361/201731320}

\bibitem[{{Aguado} {et~al.}(2017{\natexlab{b}}){Aguado}, {Gonz{\'a}lez
  Hern{\'a}ndez}, {Allende Prieto}, \& {Rebolo}}]{Aguado2017b}
{Aguado}, D.~S., {Gonz{\'a}lez Hern{\'a}ndez}, J.~I., {Allende Prieto}, C., \&
  {Rebolo}, R. 2017{\natexlab{b}}, \aap, 605, A40,
  \dodoi{10.1051/0004-6361/201730654}

\bibitem[{{Aguado} {et~al.}(2018{\natexlab{b}}){Aguado}, {Gonz{\'a}lez
  Hern{\'a}ndez}, {Allende Prieto}, \& {Rebolo}}]{Aguado2018a}
---. 2018{\natexlab{b}}, \apjl, 852, L20, \dodoi{10.3847/2041-8213/aaa23a}

\bibitem[{Ahumada {et~al.}(2020)Ahumada, Prieto, Almeida, Anders, Anderson,
  Andrews, Anguiano, Arcodia, Armengaud, Aubert, Avila, Avila-Reese, Badenes,
  Balland, Barger, Barrera-Ballesteros, Basu, Bautista, Beaton, Beers,
  Benavides, Bender, Bernardi, Bershady, Beutler, Bidin, Bird, Bizyaev, Blanc,
  Blanton, Boquien, Borissova, Bovy, Brandt, Brinkmann, Brownstein, Bundy,
  Bureau, Burgasser, Burtin, Cano-D{\'{i}}az, Capasso, Cappellari, Carrera,
  Chabanier, Chaplin, Chapman, Cherinka, Chiappini, {Doohyun Choi}, Chojnowski,
  Chung, Clerc, Coffey, Comerford, Comparat, da~Costa, Cousinou, Covey, Crane,
  Cunha, Ilha, Dai, Damsted, Darling, Davidson, Davies, Dawson, De, de~la
  Macorra, {De Lee}, Queiroz, {Deconto Machado}, de~la Torre, Dell'Agli, {du
  Mas des Bourboux}, Diamond-Stanic, Dillon, Donor, Drory, Duckworth, Dwelly,
  Ebelke, Eftekharzadeh, {Davis Eigenbrot}, Elsworth, Eracleous, Erfanianfar,
  Escoffier, Fan, Farr, Fern{\'{a}}ndez-Trincado, Feuillet, Finoguenov, Fofie,
  Fraser-McKelvie, Frinchaboy, Fromenteau, Fu, Galbany, Garcia,
  Garc{\'{i}}a-Hern{\'{a}}ndez, Oehmichen, Ge, Maia, Geisler, Gelfand, Goddy,
  Gonzalez-Perez, Grabowski, Green, Grier, Guo, Guy, Harding, Hasselquist,
  Hawken, Hayes, Hearty, Hekker, Hogg, Holtzman, Horta, Hou, Hsieh, Huber,
  Hunt, Chitham, Imig, Jaber, Angel, Johnson, Jones, J{\"{o}}nsson, Jullo, Kim,
  Kinemuchi, {Kirkpatrick IV}, Kite, Klaene, Kneib, Kollmeier, Kong, Kounkel,
  Krishnarao, Lacerna, Lan, Lane, Law, {Le Goff}, Leung, Lewis, Li, Lian, Lin,
  Long, Longa-Pe{\~{n}}a, Lundgren, Lyke, {Ted Mackereth}, MacLeod, Majewski,
  Manchado, Maraston, Martini, Masseron, Masters, Mathur, McDermid, Merloni,
  Merrifield, M{\'{e}}sz{\'{a}}ros, Miglio, Minniti, Minsley, Miyaji, Mohammad,
  Mosser, Mueller, Muna, Mu{\~{n}}oz-Guti{\'{e}}rrez, Myers, Nadathur, Nair,
  Nandra, do~Nascimento, Nevin, Newman, Nidever, Nitschelm, Noterdaeme,
  O'Connell, Olmstead, Oravetz, Oravetz, Osorio, Pace, Padilla,
  Palanque-Delabrouille, Palicio, Pan, Pan, Parker, Paviot, Peirani, Ramŕez,
  Penny, Percival, Perez-Fournon, P{\'{e}}rez-R{\`{a}}fols, Petitjean, Pieri,
  Pinsonneault, Poovelil, Povick, Prakash, Price-Whelan, Raddick, Raichoor,
  Ray, Rembold, Rezaie, Riffel, Riffel, Rix, Robin, Roman-Lopes,
  Rom{\'{a}}n-Z{\'{u}}{\~{n}}iga, Rose, Ross, Rossi, Rowlands, Rubin, Salvato,
  S{\'{a}}nchez, S{\'{a}}nchez-Menguiano, S{\'{a}}nchez-Gallego, Sayres,
  Schaefer, Schiavon, Schimoia, Schlafly, Schlegel, Schneider, Schultheis,
  Schwope, Seo, Serenelli, Shafieloo, Shamsi, Shao, Shen, Shetrone, Shirley,
  Aguirre, Simon, Skrutskie, Slosar, Smethurst, Sobeck, Sodi, Souto, Stark,
  Stassun, Steinmetz, Stello, Stermer, Storchi-Bergmann, Streblyanska,
  Stringfellow, Stutz, Su{\'{a}}rez, Sun, Taghizadeh-Popp, Talbot, Tayar,
  Thakar, Theriault, Thomas, Thomas, Tinker, Tojeiro, Toledo, Tremonti, Troup,
  Tuttle, Unda-Sanzana, Valentini, Vargas-Gonz{\'{a}}lez, Vargas-Maga{\~{n}}a,
  V{\'{a}}zquez-Mata, Vivek, Wake, Wang, Weaver, Weijmans, Wild, Wilson,
  Wilson, Wolthuis, Wood-Vasey, Yan, Yang, Y{\`{e}}che, Zamora, Zarrouk,
  Zasowski, Zhang, Zhao, Zhao, Zheng, Zheng, Zhu, \& Zou}]{Ahumada2020}
Ahumada, R., Prieto, C.~A., Almeida, A., {et~al.} 2020, The Astrophysical
  Journal Supplement Series, 249, 3, \dodoi{10.3847/1538-4365/ab929e}

\bibitem[{{Allende Prieto} {et~al.}(2006){Allende Prieto}, {Beers}, {Wilhelm},
  {Newberg}, {Rockosi}, {Yanny}, \& {Lee}}]{Allende_Prieto2006}
{Allende Prieto}, C., {Beers}, T.~C., {Wilhelm}, R., {et~al.} 2006, \apj, 636,
  804, \dodoi{10.1086/498131}

\bibitem[{{Allende Prieto} {et~al.}(2008){Allende Prieto}, {Sivarani}, {Beers},
  {Lee}, {Koesterke}, {Shetrone}, {Sneden}, {Lambert}, {Wilhelm}, {Rockosi},
  {Lai}, {Yanny}, {Ivans}, {Johnson}, {Aoki}, {Bailer-Jones}, \& {Re
  Fiorentin}}]{Allende_Prieto2008}
{Allende Prieto}, C., {Sivarani}, T., {Beers}, T.~C., {et~al.} 2008, \aj, 136,
  2070, \dodoi{10.1088/0004-6256/136/5/2070}

\bibitem[{{Allende Prieto} {et~al.}(2014){Allende Prieto},
  {Fern{\'a}ndez-Alvar}, {Schlesinger}, {Lee}, {Morrison}, {Schneider},
  {Beers}, {Bizyaev}, {Ebelke}, {Malanushenko}, {Malanushenko}, {Oravetz},
  {Pan}, {Simmons}, {Simmerer}, {Sobeck}, \& {Robin}}]{Allende_Prieto2014}
{Allende Prieto}, C., {Fern{\'a}ndez-Alvar}, E., {Schlesinger}, K.~J., {et~al.}
  2014, \aap, 568, A7, \dodoi{10.1051/0004-6361/201424053}

\bibitem[{{Allende Prieto} {et~al.}(2015){Allende Prieto},
  {Fern{\'a}ndez-Alvar}, {Aguado}, {Gonz{\'a}lez Hern{\'a}ndez}, {Rebolo},
  {Lee}, {Beers}, {Rockosi}, \& {Ge}}]{Allende_Prieto2015}
{Allende Prieto}, C., {Fern{\'a}ndez-Alvar}, E., {Aguado}, D.~S., {et~al.}
  2015, \aap, 579, A98, \dodoi{10.1051/0004-6361/201525904}

\bibitem[{Aoki {et~al.}(2013)Aoki, Beers, Lee, Honda, Ito, Takada-Hidai,
  Frebel, Suda, Fujimoto, Carollo, \& Sivarani}]{Aoki2013}
Aoki, W., Beers, T.~C., Lee, Y.~S., {et~al.} 2013, Astronomical Journal, 145,
  \dodoi{10.1088/0004-6256/145/1/13}

\bibitem[{{Arenou} {et~al.}(2018){Arenou}, {Luri}, {Babusiaux}, {Fabricius},
  {Helmi}, {Muraveva}, {Robin}, {Spoto}, {Vallenari}, {Antoja},
  {Cantat-Gaudin}, {Jordi}, {Leclerc}, {Reyl{\'e}}, {Romero-G{\'o}mez}, {Shih},
  {Soria}, {Barache}, {Bossini}, {Bragaglia}, {Breddels}, {Fabrizio},
  {Lambert}, {Marrese}, {Massari}, {Moitinho}, {Robichon}, {Ruiz-Dern},
  {Sordo}, {Veljanoski}, {Eyer}, {Jasniewicz}, {Pancino}, {Soubiran}, {Spagna},
  {Tanga}, {Turon}, \& {Zurbach}}]{Arenou2018}
{Arenou}, F., {Luri}, X., {Babusiaux}, C., {et~al.} 2018, \aap, 616, A17,
  \dodoi{10.1051/0004-6361/201833234}

\bibitem[{{Arentsen} {et~al.}(2019){Arentsen}, {Starkenburg}, {Shetrone},
  {Venn}, {Depagne}, \& {McConnachie}}]{Arentsen2019}
{Arentsen}, A., {Starkenburg}, E., {Shetrone}, M.~D., {et~al.} 2019, A\&A, 621,
  A108, \dodoi{10.1051/0004-6361/201834146}

\bibitem[{{Bailer-Jones}(2015)}]{Bailer-Jones2015}
{Bailer-Jones}, C. A.~L. 2015, PASP, 127, 994, \dodoi{10.1086/683116}

\bibitem[{{Beers} \& {Christlieb}(2005)}]{2005ARA&A..43..531B}
{Beers}, T.~C., \& {Christlieb}, N. 2005, \araa, 43, 531,
  \dodoi{10.1146/annurev.astro.42.053102.134057}

\bibitem[{{Bessell} {et~al.}(2011){Bessell}, {Bloxham}, {Schmidt}, {Keller},
  {Tisserand}, \& {Francis}}]{Bessell2011}
{Bessell}, M., {Bloxham}, G., {Schmidt}, B., {et~al.} 2011, \pasp, 123, 789,
  \dodoi{10.1086/660849}

\bibitem[{Blanton {et~al.}(2017)Blanton, Bershady, Abolfathi, Albareti, Prieto,
  Almeida, Alonso-Garc{\'{i}}a, Anders, Anderson, Andrews, Aquino-Ort{\'{i}}z,
  Arag{\'{o}}n-Salamanca, Argudo-Fern{\'{a}}ndez, Armengaud, Aubourg,
  Avila-Reese, Badenes, Bailey, Barger, Barrera-Ballesteros, Bartosz, Bates,
  Baumgarten, Bautista, Beaton, Beers, Belfiore, Bender, Berlind, Bernardi,
  Beutler, Bird, Bizyaev, Blanc, Blomqvist, Bolton, Boquien, Borissova, van~den
  Bosch, Bovy, Brandt, Brinkmann, Brownstein, Bundy, Burgasser, Burtin, Busca,
  Cappellari, Carigi, Carlberg, Rosell, Carrera, Chanover, Cherinka, Cheung,
  Chew, Chiappini, Choi, Chojnowski, Chuang, Chung, Cirolini, Clerc, Cohen,
  Comparat, da~Costa, Cousinou, Covey, Crane, Croft, Cruz-Gonzalez, Cuadra,
  Cunha, Damke, Darling, Davies, Dawson, de~la Macorra, Dell'Agli, Lee,
  Delubac, Mille, Diamond-Stanic, Cano-D{\'{i}}az, Donor, Downes, Drory, des
  Bourboux, Duckworth, Dwelly, Dyer, Ebelke, Eigenbrot, Eisenstein, Emsellem,
  Eracleous, Escoffier, Evans, Fan, Fern{\'{a}}ndez-Alvar, Fernandez-Trincado,
  Feuillet, Finoguenov, Fleming, Font-Ribera, Fredrickson, Freischlad,
  Frinchaboy, Fuentes, Galbany, Garcia-Dias, Garc{\'{i}}a-Hern{\'{a}}ndez,
  Gaulme, Geisler, Gelfand, Gil-Mar{\'{i}}n, Gillespie, Goddard,
  Gonzalez-Perez, Grabowski, Green, Grier, Gunn, Guo, Guy, Hagen, Hahn, Hall,
  Harding, Hasselquist, Hawley, Hearty, Hern{\'{a}}ndez, Ho, Hogg,
  Holley-Bockelmann, Holtzman, Holzer, Huehnerhoff, Hutchinson, Hwang,
  Ibarra-Medel, Ilha, Ivans, Ivory, Jackson, Jensen, Johnson, Jones,
  J{\"{o}}nsson, Jullo, Kamble, Kinemuchi, Kirkby, Kitaura, Klaene, Knapp,
  Kneib, Kollmeier, Lacerna, Lane, Lang, Law, Lazarz, Lee, Goff, Liang, Li, Li,
  Lian, Lima, Lin, Lin, de~Lis, Liu, Lizaola, Long, Lucatello, Lundgren,
  MacDonald, Machado, MacLeod, Mahadevan, Maia, Maiolino, Majewski,
  Malanushenko, Malanushenko, Manchado, Mao, Maraston, Marques-Chaves,
  Masseron, Masters, McBride, McDermid, McGrath, McGreer, {Medina Pe{\~{n}}a},
  Melendez, Merloni, Merrifield, Meszaros, Meza, Minchev, Minniti, Miyaji,
  More, Mulchaey, M{\"{u}}ller-S{\'{a}}nchez, Muna, Munoz, Myers, Nair, Nandra,
  do~Nascimento, Negrete, Ness, Newman, Nichol, Nidever, Nitschelm, Ntelis,
  O'Connell, Oelkers, Oravetz, Oravetz, Pace, Padilla, Palanque-Delabrouille,
  Palicio, Pan, Parejko, Parikh, P{\^{a}}ris, Park, Patten, Peirani,
  Pellejero-Ibanez, Penny, Percival, Perez-Fournon, Petitjean, Pieri,
  Pinsonneault, Pisani, Poleski, Prada, Prakash, Queiroz, Raddick, Raichoor,
  Rembold, Richstein, Riffel, Riffel, Rix, Robin, Rockosi,
  Rodr{\'{i}}guez-Torres, Roman-Lopes, Rom{\'{a}}n-Z{\'{u}}{\~{n}}iga, Rosado,
  Ross, Rossi, Ruan, Ruggeri, Rykoff, Salazar-Albornoz, Salvato, S{\'{a}}nchez,
  Aguado, S{\'{a}}nchez-Gallego, Santana, Santiago, Sayres, Schiavon, Schimoia,
  Schlafly, Schlegel, Schneider, Schultheis, Schuster, Schwope, Seo, Shao,
  Shen, Shetrone, Shull, Simon, Skinner, Skrutskie, Slosar, Smith, Sobeck,
  Sobreira, Somers, Souto, Stark, Stassun, Stauffer, Steinmetz,
  Storchi-Bergmann, Streblyanska, Stringfellow, Su{\'{a}}rez, Sun, Suzuki,
  Szigeti, Taghizadeh-Popp, Tang, Tao, Tayar, Tembe, Teske, Thakar, Thomas,
  Thompson, Tinker, Tissera, Tojeiro, Toledo, de~la Torre, Tremonti, Troup,
  Valenzuela, Valpuesta, Vargas-Gonz{\'{a}}lez, Vargas-Maga{\~{n}}a, Vazquez,
  Villanova, Vivek, Vogt, Wake, Walterbos, Wang, Weaver, Weijmans, Weinberg,
  Westfall, Whelan, Wild, Wilson, Wood-Vasey, Wylezalek, Xiao, Yan, Yang,
  Ybarra, Y{\`{e}}che, Zakamska, Zamora, Zarrouk, Zasowski, Zhang, Zhao, Zheng,
  Zheng, Zhou, Zhou, Zhu, Zoccali, \& Zou}]{Blanton2017}
Blanton, M.~R., Bershady, M.~A., Abolfathi, B., {et~al.} 2017, The Astronomical
  Journal, 154, 28, \dodoi{10.3847/1538-3881/aa7567}

\bibitem[{Blouin {et~al.}(2018{\natexlab{a}})Blouin, Dufour, \&
  Allard}]{Blouin2018}
Blouin, S., Dufour, P., \& Allard, N.~F. 2018{\natexlab{a}}, The Astrophysical
  Journal, 863, 184, \dodoi{10.3847/1538-4357/aad4a9}

\bibitem[{Blouin {et~al.}(2018{\natexlab{b}})Blouin, Dufour, Allard, \&
  Kilic}]{Blouin2018a}
Blouin, S., Dufour, P., Allard, N.~F., \& Kilic, M. 2018{\natexlab{b}}, The
  Astrophysical Journal, 867, 161, \dodoi{10.3847/1538-4357/aae53a}

\bibitem[{Blouin {et~al.}(2019)Blouin, Dufour, Thibeault, \&
  Allard}]{Blouin2019}
Blouin, S., Dufour, P., Thibeault, C., \& Allard, N.~F. 2019, The Astrophysical
  Journal, 878, 63, \dodoi{10.3847/1538-4357/ab1f82}

\bibitem[{Blouin {et~al.}(2017)Blouin, Kowalski, \& Dufour}]{Blouin2017}
Blouin, S., Kowalski, P.~M., \& Dufour, P. 2017, The Astrophysical Journal,
  848, 36, \dodoi{10.3847/1538-4357/aa8ad6}

\bibitem[{{Bonifacio} {et~al.}(2012){Bonifacio}, Sbordone, Caffau, Ludwig,
  Spite, {Gonz{\'{a}}lez Hern{\'{a}}ndez}, \& Behara}]{Bonifacio2012}
{Bonifacio}, P., Sbordone, L., Caffau, E., {et~al.} 2012, Astronomy and
  Astrophysics, 542, 1, \dodoi{10.1051/0004-6361/201219004}

\bibitem[{{Bonifacio} {et~al.}(2015){Bonifacio}, {Caffau}, {Spite}, {Limongi},
  {Chieffi}, {Klessen}, {Fran{\c{c}}ois}, {Molaro}, {Ludwig}, {Zaggia},
  {Spite}, {Plez}, {Cayrel}, {Christlieb}, {Clark}, {Glover}, {Hammer}, {Koch},
  {Monaco}, {Sbordone}, \& {Steffen}}]{Bonifacio2015}
{Bonifacio}, P., {Caffau}, E., {Spite}, M., {et~al.} 2015, \aap, 579, A28,
  \dodoi{10.1051/0004-6361/201425266}

\bibitem[{{Bonifacio} {et~al.}(2018){Bonifacio}, {Caffau}, {Spite}, {Spite},
  {Sbordone}, {Monaco}, {Fran{\c{c}}ois}, {Plez}, {Molaro}, {Gallagher},
  {Cayrel}, {Christlieb}, {Klessen}, {Koch}, {Ludwig}, {Steffen}, {Zaggia}, \&
  {Abate}}]{Bonifacio2018}
---. 2018, A\&A, 612, A65, \dodoi{10.1051/0004-6361/201732320}

\bibitem[{Bovy {et~al.}(2016)Bovy, Rix, Green, Schlafly, \&
  Finkbeiner}]{Bovy2016}
Bovy, J., Rix, H.-w., Green, G.~M., Schlafly, E.~F., \& Finkbeiner, D.~P. 2016,
  The Astrophysical Journal, 818, 130, \dodoi{10.3847/0004-637x/818/2/130}

\bibitem[{Bressan {et~al.}(2012)Bressan, Marigo, Girardi, Salasnich, {Dal
  Cero}, Rubele, \& Nanni}]{Bressan2012}
Bressan, A., Marigo, P., Girardi, L., {et~al.} 2012, Monthly Notices of the
  Royal Astronomical Society, 427, 127,
  \dodoi{10.1111/j.1365-2966.2012.21948.x}

\bibitem[{{Bromm}(2013)}]{Bromm2013}
{Bromm}, V. 2013, Reports on Progress in Physics, 76, 112901,
  \dodoi{10.1088/0034-4885/76/11/112901}

\bibitem[{{Bromm} {et~al.}(1999){Bromm}, {Coppi}, \& {Larson}}]{Bromm1999}
{Bromm}, V., {Coppi}, P.~S., \& {Larson}, R.~B. 1999, \apjl, 527, L5,
  \dodoi{10.1086/312385}

\bibitem[{{Bromm} {et~al.}(2002){Bromm}, {Coppi}, \& {Larson}}]{Bromm2002}
---. 2002, \apj, 564, 23, \dodoi{10.1086/323947}

\bibitem[{Brown {et~al.}(2010)Brown, Sahu, Anderson, Tumlinson, Valenti, Smith,
  Jeffery, Renzini, Zoccali, Ferguson, VandenBerg, Bond, Casertano, Valenti,
  Minniti, Livio, \& Panagia}]{Brown2010}
Brown, T.~M., Sahu, K., Anderson, J., {et~al.} 2010, Astrophysical Journal
  Letters, 725, 19, \dodoi{10.1088/2041-8205/725/1/L19}

\bibitem[{Brown {et~al.}(2017)Brown, Kilic, \& Gianninas}]{Brown2017}
Brown, W.~R., Kilic, M., \& Gianninas, A. 2017, The Astrophysical Journal, 839,
  23, \dodoi{10.3847/1538-4357/aa67e4}

\bibitem[{{Burrows} {et~al.}(1993){Burrows}, {Hubbard}, {Saumon}, \&
  {Lunine}}]{Burrows1993}
{Burrows}, A., {Hubbard}, W.~B., {Saumon}, D., \& {Lunine}, J.~I. 1993, \apj,
  406, 158, \dodoi{10.1086/172427}

\bibitem[{{Caffau} {et~al.}(2011{\natexlab{a}}){Caffau}, {Bonifacio},
  {Fran{\c{c}}ois}, Sbordone, Monaco, Spite, Spite, Ludwig, Cayrel, Zaggia,
  Hammer, Randich, Molaro, \& Hill}]{Caffau2011}
{Caffau}, E., {Bonifacio}, P., {Fran{\c{c}}ois}, P., {et~al.}
  2011{\natexlab{a}}, Nature, 477, 67, \dodoi{10.1038/nature10377}

\bibitem[{{Caffau} {et~al.}(2011{\natexlab{b}}){Caffau}, {Bonifacio},
  {Fran{\c{c}}ois}, {Spite}, {Spite}, {Zaggia}, {Ludwig}, {Monaco}, {Sbordone},
  {Cayrel}, {Hammer}, {Randich}, {Hill}, \& {Molaro}}]{Caffau2011b}
---. 2011{\natexlab{b}}, \aap, 534, A4, \dodoi{10.1051/0004-6361/201117530}

\bibitem[{{Caffau} {et~al.}(2013){Caffau}, {Bonifacio}, {Fran{\c{c}}ois},
  {Sbordone}, {Spite}, {Monaco}, {Plez}, {Spite}, {Zaggia}, {Ludwig}, {Cayrel},
  {Molaro}, {Randich}, {Hammer}, \& {Hill}}]{Caffau2013}
---. 2013, A\&A, 560, A15, \dodoi{10.1051/0004-6361/201322213}

\bibitem[{{Carbon} {et~al.}(2017){Carbon}, {Henze}, \& {Nelson}}]{Carbon2017}
{Carbon}, D.~F., {Henze}, C., \& {Nelson}, B.~C. 2017, \apjs, 228, 19,
  \dodoi{10.3847/1538-4365/228/2/19}

\bibitem[{{Chambers} {et~al.}(2016){Chambers}, {Magnier}, {Metcalfe},
  {Flewelling}, {Huber}, {Waters}, {Denneau}, {Draper}, {Farrow}, {Finkbeiner},
  {Holmberg}, {Koppenhoefer}, {Price}, {Rest}, {Saglia}, {Schlafly}, {Smartt},
  {Sweeney}, {Wainscoat}, {Burgett}, {Chastel}, {Grav}, {Heasley}, {Hodapp},
  {Jedicke}, {Kaiser}, {Kudritzki}, {Luppino}, {Lupton}, {Monet}, {Morgan},
  {Onaka}, {Shiao}, {Stubbs}, {Tonry}, {White}, {Ba{\~n}ados}, {Bell},
  {Bender}, {Bernard}, {Boegner}, {Boffi}, {Botticella}, {Calamida},
  {Casertano}, {Chen}, {Chen}, {Cole}, {Deacon}, {Frenk}, {Fitzsimmons},
  {Gezari}, {Gibbs}, {Goessl}, {Goggia}, {Gourgue}, {Goldman}, {Grant},
  {Grebel}, {Hambly}, {Hasinger}, {Heavens}, {Heckman}, {Henderson}, {Henning},
  {Holman}, {Hopp}, {Ip}, {Isani}, {Jackson}, {Keyes}, {Koekemoer}, {Kotak},
  {Le}, {Liska}, {Long}, {Lucey}, {Liu}, {Martin}, {Masci}, {McLean}, {Mindel},
  {Misra}, {Morganson}, {Murphy}, {Obaika}, {Narayan}, {Nieto-Santisteban},
  {Norberg}, {Peacock}, {Pier}, {Postman}, {Primak}, {Rae}, {Rai}, {Riess},
  {Riffeser}, {Rix}, {R{\"o}ser}, {Russel}, {Rutz}, {Schilbach}, {Schultz},
  {Scolnic}, {Strolger}, {Szalay}, {Seitz}, {Small}, {Smith}, {Soderblom},
  {Taylor}, {Thomson}, {Taylor}, {Thakar}, {Thiel}, {Thilker}, {Unger},
  {Urata}, {Valenti}, {Wagner}, {Walder}, {Walter}, {Watters}, {Werner},
  {Wood-Vasey}, \& {Wyse}}]{Chambers2016}
{Chambers}, K.~C., {Magnier}, E.~A., {Metcalfe}, N., {et~al.} 2016, arXiv
  e-prints, arXiv:1612.05560.
\newblock \doarXiv{1612.05560}

\bibitem[{{Chen} {et~al.}(2015){Chen}, {Bressan}, {Girardi}, {Marigo}, {Kong},
  \& {Lanza}}]{Chen2015}
{Chen}, Y., {Bressan}, A., {Girardi}, L., {et~al.} 2015, \mnras, 452, 1068,
  \dodoi{10.1093/mnras/stv1281}

\bibitem[{{Chen} {et~al.}(2014){Chen}, {Girardi}, {Bressan}, {Marigo},
  {Barbieri}, \& {Kong}}]{Chen2014}
{Chen}, Y., {Girardi}, L., {Bressan}, A., {et~al.} 2014, \mnras, 444, 2525,
  \dodoi{10.1093/mnras/stu1605}

\bibitem[{{Choi} {et~al.}(2016){Choi}, {Dotter}, {Conroy}, {Cantiello},
  {Paxton}, \& {Johnson}}]{Choi2016}
{Choi}, J., {Dotter}, A., {Conroy}, C., {et~al.} 2016, ApJ, 823, 102,
  \dodoi{10.3847/0004-637X/823/2/102}

\bibitem[{{Clark} {et~al.}(2011{\natexlab{a}}){Clark}, {Glover}, {Klessen}, \&
  {Bromm}}]{Clark2011a}
{Clark}, P.~C., {Glover}, S. C.~O., {Klessen}, R.~S., \& {Bromm}, V.
  2011{\natexlab{a}}, \apj, 727, 110, \dodoi{10.1088/0004-637X/727/2/110}

\bibitem[{{Clark} {et~al.}(2011{\natexlab{b}}){Clark}, {Glover}, {Smith},
  {Greif}, {Klessen}, \& {Bromm}}]{Clark2011b}
{Clark}, P.~C., {Glover}, S. C.~O., {Smith}, R.~J., {et~al.}
  2011{\natexlab{b}}, Science, 331, 1040, \dodoi{10.1126/science.1198027}

\bibitem[{{Cohen} {et~al.}(2013){Cohen}, {Christlieb}, {Thompson}, {McWilliam},
  {Shectman}, {Reimers}, {Wisotzki}, \& {Kirby}}]{2013ApJ...778...56C}
{Cohen}, J.~G., {Christlieb}, N., {Thompson}, I., {et~al.} 2013, \apj, 778, 56,
  \dodoi{10.1088/0004-637X/778/1/56}

\bibitem[{{Cui} {et~al.}(2012){Cui}, {Zhao}, {Chu}, {Li}, {Li}, {Zhang}, {Su},
  {Yao}, {Wang}, {Xing}, {Li}, {Zhu}, {Wang}, {Gu}, {Luo}, {Xu}, {Zhang},
  {Liu}, {Zhang}, {Yang}, {Cao}, {Chen}, {Chen}, {Chen}, {Chen}, {Chu}, {Feng},
  {Gong}, {Hou}, {Hu}, {Hu}, {Hu}, {Jia}, {Jiang}, {Jiang}, {Jiang}, {Jin},
  {Li}, {Li}, {Li}, {Liu}, {Liu}, {Lu}, {Mao}, {Men}, {Qi}, {Qi}, {Shi},
  {Tang}, {Tao}, {Wang}, {Wang}, {Wang}, {Wang}, {Wang}, {Wang}, {Wang},
  {Wang}, {Wang}, {Wang}, {Wang}, {Wang}, {Xu}, {Xu}, {Yang}, {Yu}, {Yuan},
  {Yuan}, {Zhai}, {Zhang}, {Zhang}, {Zhang}, {Zhao}, {Zhou}, {Zhou}, {Zhu}, \&
  {Zou}}]{2012RAA....12.1197C}
{Cui}, X.-Q., {Zhao}, Y.-H., {Chu}, Y.-Q., {et~al.} 2012, Research in Astronomy
  and Astrophysics, 12, 1197, \dodoi{10.1088/1674-4527/12/9/003}

\bibitem[{{Dalton} {et~al.}(2012){Dalton}, {Trager}, {Abrams}, {Carter},
  {Bonifacio}, {Aguerri}, {MacIntosh}, {Evans}, {Lewis}, {Navarro}, {Agocs},
  {Dee}, {Rousset}, {Tosh}, {Middleton}, {Pragt}, {Terrett}, {Brock}, {Benn},
  {Verheijen}, {Cano Infantes}, {Bevil}, {Steele}, {Mottram}, {Bates},
  {Gribbin}, {Rey}, {Rodriguez}, {Delgado}, {Guinouard}, {Walton}, {Irwin},
  {Jagourel}, {Stuik}, {Gerlofsma}, {Roelfsma}, {Skillen}, {Ridings},
  {Balcells}, {Daban}, {Gouvret}, {Venema}, \& {Girard}}]{WEAVE}
{Dalton}, G., {Trager}, S.~C., {Abrams}, D.~C., {et~al.} 2012, in Society of
  Photo-Optical Instrumentation Engineers (SPIE) Conference Series, Vol. 8446,
  Ground-based and Airborne Instrumentation for Astronomy IV, 84460P,
  \dodoi{10.1117/12.925950}

\bibitem[{{Dawson} {et~al.}(2013){Dawson}, {Schlegel}, {Ahn}, {Anderson},
  {Aubourg}, {Bailey}, {Barkhouser}, {Bautista}, {Beifiori}, {Berlind},
  {Bhardwaj}, {Bizyaev}, {Blake}, {Blanton}, {Blomqvist}, {Bolton}, {Borde},
  {Bovy}, {Brandt}, {Brewington}, {Brinkmann}, {Brown}, {Brownstein}, {Bundy},
  {Busca}, {Carithers}, {Carnero}, {Carr}, {Chen}, {Comparat}, {Connolly},
  {Cope}, {Croft}, {Cuesta}, {da Costa}, {Davenport}, {Delubac}, {de Putter},
  {Dhital}, {Ealet}, {Ebelke}, {Eisenstein}, {Escoffier}, {Fan}, {Filiz Ak},
  {Finley}, {Font-Ribera}, {G{\'e}nova-Santos}, {Gunn}, {Guo}, {Haggard},
  {Hall}, {Hamilton}, {Harris}, {Harris}, {Ho}, {Hogg}, {Holder}, {Honscheid},
  {Huehnerhoff}, {Jordan}, {Jordan}, {Kauffmann}, {Kazin}, {Kirkby}, {Klaene},
  {Kneib}, {Le Goff}, {Lee}, {Long}, {Loomis}, {Lundgren}, {Lupton}, {Maia},
  {Makler}, {Malanushenko}, {Malanushenko}, {Mandelbaum}, {Manera}, {Maraston},
  {Margala}, {Masters}, {McBride}, {McDonald}, {McGreer}, {McMahon}, {Mena},
  {Miralda-Escud{\'e}}, {Montero-Dorta}, {Montesano}, {Muna}, {Myers},
  {Naugle}, {Nichol}, {Noterdaeme}, {Nuza}, {Olmstead}, {Oravetz}, {Oravetz},
  {Owen}, {Padmanabhan}, {Palanque-Delabrouille}, {Pan}, {Parejko},
  {P{\^a}ris}, {Percival}, {P{\'e}rez-Fournon}, {P{\'e}rez-R{\`a}fols},
  {Petitjean}, {Pfaffenberger}, {Pforr}, {Pieri}, {Prada}, {Price-Whelan},
  {Raddick}, {Rebolo}, {Rich}, {Richards}, {Rockosi}, {Roe}, {Ross}, {Ross},
  {Rossi}, {Rubi{\~n}o-Martin}, {Samushia}, {S{\'a}nchez}, {Sayres}, {Schmidt},
  {Schneider}, {Sc{\'o}ccola}, {Seo}, {Shelden}, {Sheldon}, {Shen}, {Shu},
  {Slosar}, {Smee}, {Snedden}, {Stauffer}, {Steele}, {Strauss}, {Streblyanska},
  {Suzuki}, {Swanson}, {Tal}, {Tanaka}, {Thomas}, {Tinker}, {Tojeiro},
  {Tremonti}, {Vargas Maga{\~n}a}, {Verde}, {Viel}, {Wake}, {Watson}, {Weaver},
  {Weinberg}, {Weiner}, {West}, {White}, {Wood-Vasey}, {Yeche}, {Zehavi},
  {Zhao}, \& {Zheng}}]{Dawson2013}
{Dawson}, K.~S., {Schlegel}, D.~J., {Ahn}, C.~P., {et~al.} 2013, \aj, 145, 10,
  \dodoi{10.1088/0004-6256/145/1/10}

\bibitem[{{Dawson} {et~al.}(2016){Dawson}, {Kneib}, {Percival}, {Alam},
  {Albareti}, {Anderson}, {Armengaud}, {Aubourg}, {Bailey}, {Bautista},
  {Berlind}, {Bershady}, {Beutler}, {Bizyaev}, {Blanton}, {Blomqvist},
  {Bolton}, {Bovy}, {Brandt}, {Brinkmann}, {Brownstein}, {Burtin}, {Busca},
  {Cai}, {Chuang}, {Clerc}, {Comparat}, {Cope}, {Croft}, {Cruz-Gonzalez}, {da
  Costa}, {Cousinou}, {Darling}, {de la Macorra}, {de la Torre}, {Delubac}, {du
  Mas des Bourboux}, {Dwelly}, {Ealet}, {Eisenstein}, {Eracleous}, {Escoffier},
  {Fan}, {Finoguenov}, {Font-Ribera}, {Frinchaboy}, {Gaulme}, {Georgakakis},
  {Green}, {Guo}, {Guy}, {Ho}, {Holder}, {Huehnerhoff}, {Hutchinson}, {Jing},
  {Jullo}, {Kamble}, {Kinemuchi}, {Kirkby}, {Kitaura}, {Klaene}, {Laher},
  {Lang}, {Laurent}, {Le Goff}, {Li}, {Liang}, {Lima}, {Lin}, {Lin}, {Lin},
  {Long}, {Lundgren}, {MacDonald}, {Geimba Maia}, {Malanushenko},
  {Malanushenko}, {Mariappan}, {McBride}, {McGreer}, {M{\'e}nard}, {Merloni},
  {Meza}, {Montero-Dorta}, {Muna}, {Myers}, {Nandra}, {Naugle}, {Newman},
  {Noterdaeme}, {Nugent}, {Ogando}, {Olmstead}, {Oravetz}, {Oravetz},
  {Padmanabhan}, {Palanque-Delabrouille}, {Pan}, {Parejko}, {P{\^a}ris},
  {Peacock}, {Petitjean}, {Pieri}, {Pisani}, {Prada}, {Prakash}, {Raichoor},
  {Reid}, {Rich}, {Ridl}, {Rodriguez-Torres}, {Carnero Rosell}, {Ross},
  {Rossi}, {Ruan}, {Salvato}, {Sayres}, {Schneider}, {Schlegel}, {Seljak},
  {Seo}, {Sesar}, {Shandera}, {Shu}, {Slosar}, {Sobreira}, {Streblyanska},
  {Suzuki}, {Taylor}, {Tao}, {Tinker}, {Tojeiro}, {Vargas-Maga{\~n}a}, {Wang},
  {Weaver}, {Weinberg}, {White}, {Wood-Vasey}, {Yeche}, {Zhai}, {Zhao}, {Zhao},
  {Zheng}, {Ben Zhu}, \& {Zou}}]{Dawson2016}
{Dawson}, K.~S., {Kneib}, J.-P., {Percival}, W.~J., {et~al.} 2016, \aj, 151,
  44, \dodoi{10.3847/0004-6256/151/2/44}

\bibitem[{{DESI Collaboration} {et~al.}(2016){DESI Collaboration}, {Aghamousa},
  {Aguilar}, {Ahlen}, {Alam}, {Allen}, {Allende Prieto}, {Annis}, {Bailey},
  {Balland}, {Ballester}, {Baltay}, {Beaufore}, {Bebek}, {Beers}, {Bell},
  {Bernal}, {Besuner}, {Beutler}, {Blake}, {Bleuler}, {Blomqvist}, {Blum},
  {Bolton}, {Briceno}, {Brooks}, {Brownstein}, {Buckley-Geer}, {Burden},
  {Burtin}, {Busca}, {Cahn}, {Cai}, {Cardiel-Sas}, {Carlberg}, {Carton},
  {Casas}, {Castand er}, {Cervantes-Cota}, {Claybaugh}, {Close}, {Coker},
  {Cole}, {Comparat}, {Cooper}, {Cousinou}, {Crocce}, {Cuby}, {Cunningham},
  {Davis}, {Dawson}, {de la Macorra}, {De Vicente}, {Delubac}, {Derwent},
  {Dey}, {Dhungana}, {Ding}, {Doel}, {Duan}, {Ealet}, {Edelstein},
  {Eftekharzadeh}, {Eisenstein}, {Elliott}, {Escoffier}, {Evatt}, {Fagrelius},
  {Fan}, {Fanning}, {Farahi}, {Farihi}, {Favole}, {Feng}, {Fernandez},
  {Findlay}, {Finkbeiner}, {Fitzpatrick}, {Flaugher}, {Flender}, {Font-Ribera},
  {Forero-Romero}, {Fosalba}, {Frenk}, {Fumagalli}, {Gaensicke}, {Gallo},
  {Garcia-Bellido}, {Gaztanaga}, {Pietro Gentile Fusillo}, {Gerard},
  {Gershkovich}, {Giannantonio}, {Gillet}, {Gonzalez-de-Rivera},
  {Gonzalez-Perez}, {Gott}, {Graur}, {Gutierrez}, {Guy}, {Habib}, {Heetderks},
  {Heetderks}, {Heitmann}, {Hellwing}, {Herrera}, {Ho}, {Holland}, {Honscheid},
  {Huff}, {Hutchinson}, {Huterer}, {Hwang}, {Illa Laguna}, {Ishikawa},
  {Jacobs}, {Jeffrey}, {Jelinsky}, {Jennings}, {Jiang}, {Jimenez}, {Johnson},
  {Joyce}, {Jullo}, {Juneau}, {Kama}, {Karcher}, {Karkar}, {Kehoe}, {Kennamer},
  {Kent}, {Kilbinger}, {Kim}, {Kirkby}, {Kisner}, {Kitanidis}, {Kneib},
  {Koposov}, {Kovacs}, {Koyama}, {Kremin}, {Kron}, {Kronig}, {Kueter-Young},
  {Lacey}, {Lafever}, {Lahav}, {Lambert}, {Lampton}, {Land riau}, {Lang},
  {Lauer}, {Le Goff}, {Le Guillou}, {Le Van Suu}, {Lee}, {Lee}, {Leitner},
  {Lesser}, {Levi}, {L'Huillier}, {Li}, {Liang}, {Lin}, {Linder}, {Loebman},
  {Luki{\'c}}, {Ma}, {MacCrann}, {Magneville}, {Makarem}, {Manera}, {Manser},
  {Marshall}, {Martini}, {Massey}, {Matheson}, {McCauley}, {McDonald},
  {McGreer}, {Meisner}, {Metcalfe}, {Miller}, {Miquel}, {Moustakas}, {Myers},
  {Naik}, {Newman}, {Nichol}, {Nicola}, {Nicolati da Costa}, {Nie}, {Niz},
  {Norberg}, {Nord}, {Norman}, {Nugent}, {O'Brien}, {Oh}, {Olsen}, {Padilla},
  {Padmanabhan}, {Padmanabhan}, {Palanque-Delabrouille}, {Palmese},
  {Pappalardo}, {P{\^a}ris}, {Park}, {Patej}, {Peacock}, {Peiris}, {Peng},
  {Percival}, {Perruchot}, {Pieri}, {Pogge}, {Pollack}, {Poppett}, {Prada},
  {Prakash}, {Probst}, {Rabinowitz}, {Raichoor}, {Ree}, {Refregier}, {Regal},
  {Reid}, {Reil}, {Rezaie}, {Rockosi}, {Roe}, {Ronayette}, {Roodman}, {Ross},
  {Ross}, {Rossi}, {Rozo}, {Ruhlmann-Kleider}, {Rykoff}, {Sabiu}, {Samushia},
  {Sanchez}, {Sanchez}, {Schlegel}, {Schneider}, {Schubnell}, {Secroun},
  {Seljak}, {Seo}, {Serrano}, {Shafieloo}, {Shan}, {Sharples}, {Sholl},
  {Shourt}, {Silber}, {Silva}, {Sirk}, {Slosar}, {Smith}, {Smoot}, {Som},
  {Song}, {Sprayberry}, {Staten}, {Stefanik}, {Tarle}, {Sien Tie}, {Tinker},
  {Tojeiro}, {Valdes}, {Valenzuela}, {Valluri}, {Vargas-Magana}, {Verde},
  {Walker}, {Wang}, {Wang}, {Weaver}, {Weaverdyck}, {Wechsler}, {Weinberg},
  {White}, {Yang}, {Yeche}, {Zhang}, {Zhao}, {Zheng}, {Zhou}, {Zhou}, {Zhu},
  {Zou}, \& {Zu}}]{2016arXiv161100036D}
{DESI Collaboration}, {Aghamousa}, A., {Aguilar}, J., {et~al.} 2016, arXiv
  e-prints, arXiv:1611.00036.
\newblock \doarXiv{1611.00036}

\bibitem[{{Doi} {et~al.}(2010){Doi}, {Tanaka}, {Fukugita}, {Gunn}, {Yasuda},
  {Ivezi{\'c}}, {Brinkmann}, {de Haars}, {Kleinman}, {Krzesinski}, \& {French
  Leger}}]{Doi2010}
{Doi}, M., {Tanaka}, M., {Fukugita}, M., {et~al.} 2010, \aj, 139, 1628,
  \dodoi{10.1088/0004-6256/139/4/1628}

\bibitem[{{Dopcke} {et~al.}(2013){Dopcke}, {Glover}, {Clark}, \&
  {Klessen}}]{Dopcke2013}
{Dopcke}, G., {Glover}, S. C.~O., {Clark}, P.~C., \& {Klessen}, R.~S. 2013,
  \apj, 766, 103, \dodoi{10.1088/0004-637X/766/2/103}

\bibitem[{{Dotter}(2016)}]{Dotter2016}
{Dotter}, A. 2016, ApJS, 222, 8, \dodoi{10.3847/0067-0049/222/1/8}

\bibitem[{Drimmel {et~al.}(2003)Drimmel, Cabrera-Lavers, \&
  L{\'{o}}pez-Corredoira}]{Drimmel2003}
Drimmel, R., Cabrera-Lavers, A., \& L{\'{o}}pez-Corredoira, M. 2003, Astronomy
  and Astrophysics, 409, 205, \dodoi{10.1051/0004-6361:20031070}

\bibitem[{{Eisenstein} {et~al.}(2011){Eisenstein}, {Weinberg}, {Agol},
  {Aihara}, {Allende Prieto}, {Anderson}, {Arns}, {Aubourg}, {Bailey},
  {Balbinot}, {Barkhouser}, {Beers}, {Berlind}, {Bickerton}, {Bizyaev},
  {Blanton}, {Bochanski}, {Bolton}, {Bosman}, {Bovy}, {Brandt}, {Breslauer},
  {Brewington}, {Brinkmann}, {Brown}, {Brownstein}, {Burger}, {Busca},
  {Campbell}, {Cargile}, {Carithers}, {Carlberg}, {Carr}, {Chang}, {Chen},
  {Chiappini}, {Comparat}, {Connolly}, {Cortes}, {Croft}, {Cunha}, {da Costa},
  {Davenport}, {Dawson}, {De Lee}, {Porto de Mello}, {de Simoni}, {Dean},
  {Dhital}, {Ealet}, {Ebelke}, {Edmondson}, {Eiting}, {Escoffier}, {Esposito},
  {Evans}, {Fan}, {Femen{\'\i}a Castell{\'a}}, {Dutra Ferreira}, {Fitzgerald},
  {Fleming}, {Font-Ribera}, {Ford}, {Frinchaboy}, {Garc{\'\i}a P{\'e}rez},
  {Gaudi}, {Ge}, {Ghezzi}, {Gillespie}, {Gilmore}, {Girardi}, {Gott}, {Gould},
  {Grebel}, {Gunn}, {Hamilton}, {Harding}, {Harris}, {Hawley}, {Hearty},
  {Hennawi}, {Gonz{\'a}lez Hern{\'a}ndez}, {Ho}, {Hogg}, {Holtzman},
  {Honscheid}, {Inada}, {Ivans}, {Jiang}, {Jiang}, {Johnson}, {Jordan},
  {Jordan}, {Kauffmann}, {Kazin}, {Kirkby}, {Klaene}, {Knapp}, {Kneib},
  {Kochanek}, {Koesterke}, {Kollmeier}, {Kron}, {Lampeitl}, {Lang}, {Lawler},
  {Le Goff}, {Lee}, {Lee}, {Leisenring}, {Lin}, {Liu}, {Long}, {Loomis},
  {Lucatello}, {Lundgren}, {Lupton}, {Ma}, {Ma}, {MacDonald}, {Mack},
  {Mahadevan}, {Maia}, {Majewski}, {Makler}, {Malanushenko}, {Malanushenko},
  {Mand elbaum}, {Maraston}, {Margala}, {Maseman}, {Masters}, {McBride},
  {McDonald}, {McGreer}, {McMahon}, {Mena Requejo}, {M{\'e}nard},
  {Miralda-Escud{\'e}}, {Morrison}, {Mullally}, {Muna}, {Murayama}, {Myers},
  {Naugle}, {Neto}, {Nguyen}, {Nichol}, {Nidever}, {O'Connell}, {Ogando},
  {Olmstead}, {Oravetz}, {Padmanabhan}, {Paegert}, {Palanque-Delabrouille},
  {Pan}, {Pandey}, {Parejko}, {P{\^a}ris}, {Pellegrini}, {Pepper}, {Percival},
  {Petitjean}, {Pfaffenberger}, {Pforr}, {Phleps}, {Pichon}, {Pieri}, {Prada},
  {Price-Whelan}, {Raddick}, {Ramos}, {Reid}, {Reyle}, {Rich}, {Richards},
  {Rieke}, {Rieke}, {Rix}, {Robin}, {Rocha-Pinto}, {Rockosi}, {Roe},
  {Rollinde}, {Ross}, {Ross}, {Rossetto}, {S{\'a}nchez}, {Santiago}, {Sayres},
  {Schiavon}, {Schlegel}, {Schlesinger}, {Schmidt}, {Schneider}, {Sellgren},
  {Shelden}, {Sheldon}, {Shetrone}, {Shu}, {Silverman}, {Simmerer}, {Simmons},
  {Sivarani}, {Skrutskie}, {Slosar}, {Smee}, {Smith}, {Snedden}, {Stassun},
  {Steele}, {Steinmetz}, {Stockett}, {Stollberg}, {Strauss}, {Szalay},
  {Tanaka}, {Thakar}, {Thomas}, {Tinker}, {Tofflemire}, {Tojeiro}, {Tremonti},
  {Vargas Maga{\~n}a}, {Verde}, {Vogt}, {Wake}, {Wan}, {Wang}, {Weaver},
  {White}, {White}, {Wilson}, {Wisniewski}, {Wood-Vasey}, {Yanny}, {Yasuda},
  {Y{\`e}che}, {York}, {Young}, {Zasowski}, {Zehavi}, \&
  {Zhao}}]{Eisenstein2011}
{Eisenstein}, D.~J., {Weinberg}, D.~H., {Agol}, E., {et~al.} 2011, \aj, 142,
  72, \dodoi{10.1088/0004-6256/142/3/72}

\bibitem[{Flaugher {et~al.}(2015)Flaugher, Diehl, Honscheid, Abbott, Alvarez,
  Angstadt, Annis, Antonik, Ballester, Beaufore, Bernstein, Bernstein, Bigelow,
  Bonati, Boprie, Brooks, Buckley-Geer, Campa, Cardiel-Sas, Castander,
  Castilla, Cease, Cela-Ruiz, Chappa, Chi, Cooper, {Da Costa}, Dede, Derylo,
  Depoy, {De Vicente}, Doel, Drlica-Wagner, Eiting, Elliott, Emes, Estrada,
  {Fausti Neto}, Finley, Flores, Frieman, Gerdes, Gladders, Gregory, Gutierrez,
  Hao, Holland, Holm, Huffman, Jackson, James, Jonas, Karcher, Karliner, Kent,
  Kessler, Kozlovsky, Kron, Kubik, Kuehn, Kuhlmann, Kuk, Lahav, Lathrop, Lee,
  Levi, Lewis, Li, Mandrichenko, Marshall, Martinez, Merritt, Miquel,
  Mu{\~{n}}oz, Neilsen, Nichol, Nord, Ogando, Olsen, Palaio, Patton, Peoples,
  Plazas, Rauch, Reil, Rheault, Roe, Rogers, Roodman, Sanchez, Scarpine,
  Schindler, Schmidt, Schmitt, Schubnell, Schultz, Schurter, Scott, Serrano,
  Shaw, Smith, Soares-Santos, Stefanik, Stuermer, Suchyta, Sypniewski, Tarle,
  Thaler, Tighe, Tran, Tucker, Walker, Wang, Watson, Weaverdyck, Wester, Woods,
  \& Yanny}]{Flaugher2015}
Flaugher, B., Diehl, H.~T., Honscheid, K., {et~al.} 2015, Astronomical Journal,
  \dodoi{10.1088/0004-6256/150/5/150}

\bibitem[{Foreman-Mackey {et~al.}(2013)Foreman-Mackey, Hogg, Lang, \&
  Goodman}]{Foreman-Mackey2013}
Foreman-Mackey, D., Hogg, D.~W., Lang, D., \& Goodman, J. 2013, Publications of
  the Astronomical Society of the Pacific, \dodoi{10.1086/670067}

\bibitem[{{Foreman-Mackey} {et~al.}(2019){Foreman-Mackey}, {Farr}, {Sinha},
  {Archibald}, {Hogg}, {Sanders}, {Zuntz}, {Williams}, {Nelson}, {de
  Val-Borro}, {Erhardt}, {Pashchenko}, \& {Pla}}]{Foreman-Mackey2019}
{Foreman-Mackey}, D., {Farr}, W., {Sinha}, M., {et~al.} 2019, JOSS, 4, 1864,
  \dodoi{10.21105/joss.01864}

\bibitem[{Fouesneau(2020)}]{pyphot}
Fouesneau, M. 2020, {pyphot -- A tool for computing photometry from spectra},
  GitHub.
\newblock \url{https://github.com/mfouesneau/pyphot}

\bibitem[{{Fran{\c{c}}ois} {et~al.}(2018){Fran{\c{c}}ois}, {Caffau},
  {Bonifacio}, {Spite}, {Spite}, {Cayrel}, {Christlieb}, {Gallagher},
  {Klessen}, {Koch}, {Ludwig}, {Monaco}, {Plez}, {Steffen}, \&
  {Zaggia}}]{Francois2018}
{Fran{\c{c}}ois}, P., {Caffau}, E., {Bonifacio}, P., {et~al.} 2018, A\&A, 620,
  A187, \dodoi{10.1051/0004-6361/201834375}

\bibitem[{{Fraser} {et~al.}(2017){Fraser}, {Casey}, {Gilmore}, {Heger}, \&
  {Chan}}]{Fraser2017}
{Fraser}, M., {Casey}, A.~R., {Gilmore}, G., {Heger}, A., \& {Chan}, C. 2017,
  \mnras, 468, 418, \dodoi{10.1093/mnras/stx480}

\bibitem[{Frebel {et~al.}(2015)Frebel, Chiti, Ji, Jacobson, \&
  Placco}]{Frebel2015a}
Frebel, A., Chiti, A., Ji, A.~P., Jacobson, H.~R., \& Placco, V.~M. 2015,
  Astrophysical Journal Letters, 810, L27, \dodoi{10.1088/2041-8205/810/2/L27}

\bibitem[{{Frebel} {et~al.}(2015){Frebel}, {Chiti}, {Ji}, {Jacobson}, \&
  {Placco}}]{Frebel2015b}
{Frebel}, A., {Chiti}, A., {Ji}, A.~P., {Jacobson}, H.~R., \& {Placco}, V.~M.
  2015, \apjl, 810, L27, \dodoi{10.1088/2041-8205/810/2/L27}

\bibitem[{{Frebel} \& {Norris}(2015)}]{2015ARA&A..53..631F}
{Frebel}, A., \& {Norris}, J.~E. 2015, \araa, 53, 631,
  \dodoi{10.1146/annurev-astro-082214-122423}

\bibitem[{{Fukugita} {et~al.}(1996){Fukugita}, {Ichikawa}, {Gunn}, {Doi},
  {Shimasaku}, \& {Schneider}}]{Fukugita1996}
{Fukugita}, M., {Ichikawa}, T., {Gunn}, J.~E., {et~al.} 1996, \aj, 111, 1748,
  \dodoi{10.1086/117915}

\bibitem[{{Gaia Collaboration} {et~al.}(2016){Gaia Collaboration}, {Prusti},
  {de Bruijne}, {Brown}, {Vallenari}, {Babusiaux}, {Bailer-Jones}, {Bastian},
  {Biermann}, {Evans}, {Eyer}, {Jansen}, {Jordi}, {Klioner}, {Lammers},
  {Lindegren}, {Luri}, {Mignard}, {Milligan}, {Panem}, {Poinsignon},
  {Pourbaix}, {Randich}, {Sarri}, {Sartoretti}, {Siddiqui}, {Soubiran},
  {Valette}, {van Leeuwen}, {Walton}, {Aerts}, {Arenou}, {Cropper}, {Drimmel},
  {H{\o}g}, {Katz}, {Lattanzi}, {O'Mullane}, {Grebel}, {Holland}, {Huc},
  {Passot}, {Bramante}, {Cacciari}, {Casta{\~n}eda}, {Chaoul}, {Cheek}, {De
  Angeli}, {Fabricius}, {Guerra}, {Hern{\'a}ndez}, {Jean-Antoine-Piccolo},
  {Masana}, {Messineo}, {Mowlavi}, {Nienartowicz}, {Ord{\'o}{\~n}ez-Blanco},
  {Panuzzo}, {Portell}, {Richards}, {Riello}, {Seabroke}, {Tanga},
  {Th{\'e}venin}, {Torra}, {Els}, {Gracia-Abril}, {Comoretto},
  {Garcia-Reinaldos}, {Lock}, {Mercier}, {Altmann}, {Andrae}, {Astraatmadja},
  {Bellas-Velidis}, {Benson}, {Berthier}, {Blomme}, {Busso}, {Carry},
  {Cellino}, {Clementini}, {Cowell}, {Creevey}, {Cuypers}, {Davidson}, {De
  Ridder}, {de Torres}, {Delchambre}, {Dell'Oro}, {Ducourant}, {Fr{\'e}mat},
  {Garc{\'\i}a-Torres}, {Gosset}, {Halbwachs}, {Hambly}, {Harrison}, {Hauser},
  {Hestroffer}, {Hodgkin}, {Huckle}, {Hutton}, {Jasniewicz}, {Jordan},
  {Kontizas}, {Korn}, {Lanzafame}, {Manteiga}, {Moitinho}, {Muinonen},
  {Osinde}, {Pancino}, {Pauwels}, {Petit}, {Recio-Blanco}, {Robin}, {Sarro},
  {Siopis}, {Smith}, {Smith}, {Sozzetti}, {Thuillot}, {van Reeven}, {Viala},
  {Abbas}, {Abreu Aramburu}, {Accart}, {Aguado}, {Allan}, {Allasia},
  {Altavilla}, {{\'A}lvarez}, {Alves}, {Anderson}, {Andrei}, {Anglada Varela},
  {Antiche}, {Antoja}, {Ant{\'o}n}, {Arcay}, {Atzei}, {Ayache}, {Bach},
  {Baker}, {Balaguer-N{\'u}{\~n}ez}, {Barache}, {Barata}, {Barbier}, {Barblan},
  {Baroni}, {Barrado y Navascu{\'e}s}, {Barros}, {Barstow}, {Becciani},
  {Bellazzini}, {Bellei}, {Bello Garc{\'\i}a}, {Belokurov}, {Bendjoya},
  {Berihuete}, {Bianchi}, {Bienaym{\'e}}, {Billebaud}, {Blagorodnova},
  {Blanco-Cuaresma}, {Boch}, {Bombrun}, {Borrachero}, {Bouquillon}, {Bourda},
  {Bouy}, {Bragaglia}, {Breddels}, {Brouillet}, {Br{\"u}semeister},
  {Bucciarelli}, {Budnik}, {Burgess}, {Burgon}, {Burlacu}, {Busonero}, {Buzzi},
  {Caffau}, {Cambras}, {Campbell}, {Cancelliere}, {Cantat-Gaudin}, {Carlucci},
  {Carrasco}, {Castellani}, {Charlot}, {Charnas}, {Charvet}, {Chassat},
  {Chiavassa}, {Clotet}, {Cocozza}, {Collins}, {Collins}, {Costigan}, {Crifo},
  {Cross}, {Crosta}, {Crowley}, {Dafonte}, {Damerdji}, {Dapergolas}, {David},
  {David}, {De Cat}, {de Felice}, {de Laverny}, {De Luise}, {De March}, {de
  Martino}, {de Souza}, {Debosscher}, {del Pozo}, {Delbo}, {Delgado},
  {Delgado}, {di Marco}, {Di Matteo}, {Diakite}, {Distefano}, {Dolding}, {Dos
  Anjos}, {Drazinos}, {Dur{\'a}n}, {Dzigan}, {Ecale}, {Edvardsson}, {Enke},
  {Erdmann}, {Escolar}, {Espina}, {Evans}, {Eynard Bontemps}, {Fabre},
  {Fabrizio}, {Faigler}, {Falc{\~a}o}, {Farr{\`a}s Casas}, {Faye}, {Federici},
  {Fedorets}, {Fern{\'a}ndez-Hern{\'a}ndez}, {Fernique}, {Fienga}, {Figueras},
  {Filippi}, {Findeisen}, {Fonti}, {Fouesneau}, {Fraile}, {Fraser}, {Fuchs},
  {Furnell}, {Gai}, {Galleti}, {Galluccio}, {Garabato}, {Garc{\'\i}a-Sedano},
  {Gar{\'e}}, {Garofalo}, {Garralda}, {Gavras}, {Gerssen}, {Geyer}, {Gilmore},
  {Girona}, {Giuffrida}, {Gomes}, {Gonz{\'a}lez-Marcos},
  {Gonz{\'a}lez-N{\'u}{\~n}ez}, {Gonz{\'a}lez-Vidal}, {Granvik}, {Guerrier},
  {Guillout}, {Guiraud}, {G{\'u}rpide}, {Guti{\'e}rrez-S{\'a}nchez}, {Guy},
  {Haigron}, {Hatzidimitriou}, {Haywood}, {Heiter}, {Helmi}, {Hobbs},
  {Hofmann}, {Holl}, {Holland }, {Hunt}, {Hypki}, {Icardi}, {Irwin}, {Jevardat
  de Fombelle}, {Jofr{\'e}}, {Jonker}, {Jorissen}, {Julbe}, {Karampelas},
  {Kochoska}, {Kohley}, {Kolenberg}, {Kontizas}, {Koposov}, {Kordopatis},
  {Koubsky}, {Kowalczyk}, {Krone-Martins}, {Kudryashova}, {Kull}, {Bachchan},
  {Lacoste-Seris}, {Lanza}, {Lavigne}, {Le Poncin-Lafitte}, {Lebreton},
  {Lebzelter}, {Leccia}, {Leclerc}, {Lecoeur-Taibi}, {Lemaitre}, {Lenhardt},
  {Leroux}, {Liao}, {Licata}, {Lindstr{\o}m}, {Lister}, {Livanou}, {Lobel},
  {L{\"o}ffler}, {L{\'o}pez}, {Lopez-Lozano}, {Lorenz}, {Loureiro},
  {MacDonald}, {Magalh{\~a}es Fernandes}, {Managau}, {Mann}, {Mantelet},
  {Marchal}, {Marchant}, {Marconi}, {Marie}, {Marinoni}, {Marrese},
  {Marschalk{\'o}}, {Marshall}, {Mart{\'\i}n-Fleitas}, {Martino}, {Mary},
  {Matijevi{\v{c}}}, {Mazeh}, {McMillan}, {Messina}, {Mestre}, {Michalik},
  {Millar}, {Miranda}, {Molina}, {Molinaro}, {Molinaro}, {Moln{\'a}r},
  {Moniez}, {Montegriffo}, {Monteiro}, {Mor}, {Mora}, {Morbidelli}, {Morel},
  {Morgenthaler}, {Morley}, {Morris}, {Mulone}, {Muraveva}, {Musella},
  {Narbonne}, {Nelemans}, {Nicastro}, {Noval}, {Ord{\'e}novic},
  {Ordieres-Mer{\'e}}, {Osborne}, {Pagani}, {Pagano}, {Pailler}, {Palacin},
  {Palaversa}, {Parsons}, {Paulsen}, {Pecoraro}, {Pedrosa}, {Pentik{\"a}inen},
  {Pereira}, {Pichon}, {Piersimoni}, {Pineau}, {Plachy}, {Plum}, {Poujoulet},
  {Pr{\v{s}}a}, {Pulone}, {Ragaini}, {Rago}, {Rambaux}, {Ramos-Lerate},
  {Ranalli}, {Rauw}, {Read}, {Regibo}, {Renk}, {Reyl{\'e}}, {Ribeiro},
  {Rimoldini}, {Ripepi}, {Riva}, {Rixon}, {Roelens}, {Romero-G{\'o}mez},
  {Rowell}, {Royer}, {Rudolph}, {Ruiz-Dern}, {Sadowski}, {Sagrist{\`a}
  Sell{\'e}s}, {Sahlmann}, {Salgado}, {Salguero}, {Sarasso}, {Savietto},
  {Schnorhk}, {Schultheis}, {Sciacca}, {Segol}, {Segovia}, {Segransan},
  {Serpell}, {Shih}, {Smareglia}, {Smart}, {Smith}, {Solano}, {Solitro},
  {Sordo}, {Soria Nieto}, {Souchay}, {Spagna}, {Spoto}, {Stampa}, {Steele},
  {Steidelm{\"u}ller}, {Stephenson}, {Stoev}, {Suess}, {S{\"u}veges}, {Surdej},
  {Szabados}, {Szegedi-Elek}, {Tapiador}, {Taris}, {Tauran}, {Taylor},
  {Teixeira}, {Terrett}, {Tingley}, {Trager}, {Turon}, {Ulla}, {Utrilla},
  {Valentini}, {van Elteren}, {Van Hemelryck}, {van Leeuwen}, {Varadi},
  {Vecchiato}, {Veljanoski}, {Via}, {Vicente}, {Vogt}, {Voss}, {Votruba},
  {Voutsinas}, {Walmsley}, {Weiler}, {Weingrill}, {Werner}, {Wevers},
  {Whitehead}, {Wyrzykowski}, {Yoldas}, {{\v{Z}}erjal}, {Zucker}, {Zurbach},
  {Zwitter}, {Alecu}, {Allen}, {Allende Prieto}, {Amorim},
  {Anglada-Escud{\'e}}, {Arsenijevic}, {Azaz}, {Balm}, {Beck}, {Bernstein},
  {Bigot}, {Bijaoui}, {Blasco}, {Bonfigli}, {Bono}, {Boudreault}, {Bressan},
  {Brown}, {Brunet}, {Bunclark}, {Buonanno}, {Butkevich}, {Carret}, {Carrion},
  {Chemin}, {Ch{\'e}reau}, {Corcione}, {Darmigny}, {de Boer}, {de Teodoro}, {de
  Zeeuw}, {Delle Luche}, {Domingues}, {Dubath}, {Fodor}, {Fr{\'e}zouls},
  {Fries}, {Fustes}, {Fyfe}, {Gallardo}, {Gallegos}, {Gardiol}, {Gebran},
  {Gomboc}, {G{\'o}mez}, {Grux}, {Gueguen}, {Heyrovsky}, {Hoar}, {Iannicola},
  {Isasi Parache}, {Janotto}, {Joliet}, {Jonckheere}, {Keil}, {Kim},
  {Klagyivik}, {Klar}, {Knude}, {Kochukhov}, {Kolka}, {Kos}, {Kutka}, {Lainey},
  {LeBouquin}, {Liu}, {Loreggia}, {Makarov}, {Marseille}, {Martayan},
  {Martinez-Rubi}, {Massart}, {Meynadier}, {Mignot}, {Munari}, {Nguyen},
  {Nordlander}, {Ocvirk}, {O'Flaherty}, {Olias Sanz}, {Ortiz}, {Osorio},
  {Oszkiewicz}, {Ouzounis}, {Palmer}, {Park}, {Pasquato}, {Peltzer}, {Peralta},
  {P{\'e}turaud}, {Pieniluoma}, {Pigozzi}, {Poels}, {Prat}, {Prod'homme},
  {Raison}, {Rebordao}, {Risquez}, {Rocca-Volmerange}, {Rosen}, {Ruiz-Fuertes},
  {Russo}, {Sembay}, {Serraller Vizcaino}, {Short}, {Siebert}, {Silva},
  {Sinachopoulos}, {Slezak}, {Soffel}, {Sosnowska}, {Strai{\v{z}}ys}, {ter
  Linden}, {Terrell}, {Theil}, {Tiede}, {Troisi}, {Tsalmantza}, {Tur},
  {Vaccari}, {Vachier}, {Valles}, {Van Hamme}, {Veltz}, {Virtanen}, {Wallut},
  {Wichmann}, {Wilkinson}, {Ziaeepour}, \& {Zschocke}}]{newGaia2016}
{Gaia Collaboration}, {Prusti}, T., {de Bruijne}, J.~H.~J., {et~al.} 2016,
  \aap, 595, A1, \dodoi{10.1051/0004-6361/201629272}

\bibitem[{{Gaia Collaboration} {et~al.}(2018){Gaia Collaboration}, {Brown},
  {Vallenari}, {Prusti}, {de Bruijne}, {Babusiaux}, {Bailer-Jones}, {Biermann},
  {Evans}, {Eyer}, {Jansen}, {Jordi}, {Klioner}, {Lammers}, {Lindegren},
  {Luri}, {Mignard}, {Panem}, {Pourbaix}, {Randich}, {Sartoretti}, {Siddiqui},
  {Soubiran}, {van Leeuwen}, {Walton}, {Arenou}, {Bastian}, {Cropper},
  {Drimmel}, {Katz}, {Lattanzi}, {Bakker}, {Cacciari}, {Casta{\~n}eda},
  {Chaoul}, {Cheek}, {De Angeli}, {Fabricius}, {Guerra}, {Holl}, {Masana},
  {Messineo}, {Mowlavi}, {Nienartowicz}, {Panuzzo}, {Portell}, {Riello},
  {Seabroke}, {Tanga}, {Th{\'e}venin}, {Gracia-Abril}, {Comoretto},
  {Garcia-Reinaldos}, {Teyssier}, {Altmann}, {Andrae}, {Audard},
  {Bellas-Velidis}, {Benson}, {Berthier}, {Blomme}, {Burgess}, {Busso},
  {Carry}, {Cellino}, {Clementini}, {Clotet}, {Creevey}, {Davidson}, {De
  Ridder}, {Delchambre}, {Dell'Oro}, {Ducourant},
  {Fern{\'a}ndez-Hern{\'a}ndez}, {Fouesneau}, {Fr{\'e}mat}, {Galluccio},
  {Garc{\'\i}a-Torres}, {Gonz{\'a}lez-N{\'u}{\~n}ez}, {Gonz{\'a}lez-Vidal},
  {Gosset}, {Guy}, {Halbwachs}, {Hambly}, {Harrison}, {Hern{\'a}ndez},
  {Hestroffer}, {Hodgkin}, {Hutton}, {Jasniewicz}, {Jean-Antoine-Piccolo},
  {Jordan}, {Korn}, {Krone-Martins}, {Lanzafame}, {Lebzelter}, {L{\"o}ffler},
  {Manteiga}, {Marrese}, {Mart{\'\i}n-Fleitas}, {Moitinho}, {Mora}, {Muinonen},
  {Osinde}, {Pancino}, {Pauwels}, {Petit}, {Recio-Blanco}, {Richards},
  {Rimoldini}, {Robin}, {Sarro}, {Siopis}, {Smith}, {Sozzetti}, {S{\"u}veges},
  {Torra}, {van Reeven}, {Abbas}, {Abreu Aramburu}, {Accart}, {Aerts},
  {Altavilla}, {{\'A}lvarez}, {Alvarez}, {Alves}, {Anderson}, {Andrei},
  {Anglada Varela}, {Antiche}, {Antoja}, {Arcay}, {Astraatmadja}, {Bach},
  {Baker}, {Balaguer-N{\'u}{\~n}ez}, {Balm}, {Barache}, {Barata}, {Barbato},
  {Barblan}, {Barklem}, {Barrado}, {Barros}, {Barstow}, {Bartholom{\'e}
  Mu{\~n}oz}, {Bassilana}, {Becciani}, {Bellazzini}, {Berihuete}, {Bertone},
  {Bianchi}, {Bienaym{\'e}}, {Blanco-Cuaresma}, {Boch}, {Boeche}, {Bombrun},
  {Borrachero}, {Bossini}, {Bouquillon}, {Bourda}, {Bragaglia}, {Bramante},
  {Breddels}, {Bressan}, {Brouillet}, {Br{\"u}semeister}, {Brugaletta},
  {Bucciarelli}, {Burlacu}, {Busonero}, {Butkevich}, {Buzzi}, {Caffau},
  {Cancelliere}, {Cannizzaro}, {Cantat-Gaudin}, {Carballo}, {Carlucci},
  {Carrasco}, {Casamiquela}, {Castellani}, {Castro-Ginard}, {Charlot},
  {Chemin}, {Chiavassa}, {Cocozza}, {Costigan}, {Cowell}, {Crifo}, {Crosta},
  {Crowley}, {Cuypers}, {Dafonte}, {Damerdji}, {Dapergolas}, {David}, {David},
  {de Laverny}, {De Luise}, {De March}, {de Martino}, {de Souza}, {de Torres},
  {Debosscher}, {del Pozo}, {Delbo}, {Delgado}, {Delgado}, {Di Matteo},
  {Diakite}, {Diener}, {Distefano}, {Dolding}, {Drazinos}, {Dur{\'a}n},
  {Edvardsson}, {Enke}, {Eriksson}, {Esquej}, {Eynard Bontemps}, {Fabre},
  {Fabrizio}, {Faigler}, {Falc{\~a}o}, {Farr{\`a}s Casas}, {Federici},
  {Fedorets}, {Fernique}, {Figueras}, {Filippi}, {Findeisen}, {Fonti},
  {Fraile}, {Fraser}, {Fr{\'e}zouls}, {Gai}, {Galleti}, {Garabato},
  {Garc{\'\i}a-Sedano}, {Garofalo}, {Garralda}, {Gavel}, {Gavras}, {Gerssen},
  {Geyer}, {Giacobbe}, {Gilmore}, {Girona}, {Giuffrida}, {Glass}, {Gomes},
  {Granvik}, {Gueguen}, {Guerrier}, {Guiraud}, {Guti{\'e}rrez-S{\'a}nchez},
  {Haigron}, {Hatzidimitriou}, {Hauser}, {Haywood}, {Heiter}, {Helmi}, {Heu},
  {Hilger}, {Hobbs}, {Hofmann}, {Holland}, {Huckle}, {Hypki}, {Icardi},
  {Jan{\ss}en}, {Jevardat de Fombelle}, {Jonker}, {Juh{\'a}sz}, {Julbe},
  {Karampelas}, {Kewley}, {Klar}, {Kochoska}, {Kohley}, {Kolenberg},
  {Kontizas}, {Kontizas}, {Koposov}, {Kordopatis}, {Kostrzewa-Rutkowska},
  {Koubsky}, {Lambert}, {Lanza}, {Lasne}, {Lavigne}, {Le Fustec}, {Le
  Poncin-Lafitte}, {Lebreton}, {Leccia}, {Leclerc}, {Lecoeur-Taibi},
  {Lenhardt}, {Leroux}, {Liao}, {Licata}, {Lindstr{\o}m}, {Lister}, {Livanou},
  {Lobel}, {L{\'o}pez}, {Managau}, {Mann}, {Mantelet}, {Marchal}, {Marchant},
  {Marconi}, {Marinoni}, {Marschalk{\'o}}, {Marshall}, {Martino}, {Marton},
  {Mary}, {Massari}, {Matijevi{\v{c}}}, {Mazeh}, {McMillan}, {Messina},
  {Michalik}, {Millar}, {Molina}, {Molinaro}, {Moln{\'a}r}, {Montegriffo},
  {Mor}, {Morbidelli}, {Morel}, {Morris}, {Mulone}, {Muraveva}, {Musella},
  {Nelemans}, {Nicastro}, {Noval}, {O'Mullane}, {Ord{\'e}novic},
  {Ord{\'o}{\~n}ez-Blanco}, {Osborne}, {Pagani}, {Pagano}, {Pailler},
  {Palacin}, {Palaversa}, {Panahi}, {Pawlak}, {Piersimoni}, {Pineau}, {Plachy},
  {Plum}, {Poggio}, {Poujoulet}, {Pr{\v{s}}a}, {Pulone}, {Racero}, {Ragaini},
  {Rambaux}, {Ramos-Lerate}, {Regibo}, {Reyl{\'e}}, {Riclet}, {Ripepi}, {Riva},
  {Rivard}, {Rixon}, {Roegiers}, {Roelens}, {Romero-G{\'o}mez}, {Rowell},
  {Royer}, {Ruiz-Dern}, {Sadowski}, {Sagrist{\`a} Sell{\'e}s}, {Sahlmann},
  {Salgado}, {Salguero}, {Sanna}, {Santana-Ros}, {Sarasso}, {Savietto},
  {Schultheis}, {Sciacca}, {Segol}, {Segovia}, {S{\'e}gransan}, {Shih},
  {Siltala}, {Silva}, {Smart}, {Smith}, {Solano}, {Solitro}, {Sordo}, {Soria
  Nieto}, {Souchay}, {Spagna}, {Spoto}, {Stampa}, {Steele},
  {Steidelm{\"u}ller}, {Stephenson}, {Stoev}, {Suess}, {Surdej}, {Szabados},
  {Szegedi-Elek}, {Tapiador}, {Taris}, {Tauran}, {Taylor}, {Teixeira},
  {Terrett}, {Teyssand ier}, {Thuillot}, {Titarenko}, {Torra Clotet}, {Turon},
  {Ulla}, {Utrilla}, {Uzzi}, {Vaillant}, {Valentini}, {Valette}, {van Elteren},
  {Van Hemelryck}, {van Leeuwen}, {Vaschetto}, {Vecchiato}, {Veljanoski},
  {Viala}, {Vicente}, {Vogt}, {von Essen}, {Voss}, {Votruba}, {Voutsinas},
  {Walmsley}, {Weiler}, {Wertz}, {Wevers}, {Wyrzykowski}, {Yoldas},
  {{\v{Z}}erjal}, {Ziaeepour}, {Zorec}, {Zschocke}, {Zucker}, {Zurbach}, \&
  {Zwitter}}]{newGaia2018}
{Gaia Collaboration}, {Brown}, A.~G.~A., {Vallenari}, A., {et~al.} 2018, \aap,
  616, A1, \dodoi{10.1051/0004-6361/201833051}

\bibitem[{{G{\"a}nsicke} {et~al.}(2012){G{\"a}nsicke}, {Koester}, {Farihi},
  {Girven}, {Parsons}, \& {Breedt}}]{Gaensicke2012}
{G{\"a}nsicke}, B.~T., {Koester}, D., {Farihi}, J., {et~al.} 2012, \mnras, 424,
  333, \dodoi{10.1111/j.1365-2966.2012.21201.x}

\bibitem[{{Glover}(2013)}]{Glover2013}
{Glover}, S. 2013, Astrophysics and Space Science Library, Vol. 396, {The First
  Stars} (Springer), 103, \dodoi{10.1007/978-3-642-32362-1_3}

\bibitem[{Green {et~al.}(2019)Green, Schla, Zucker, Speagle, \&
  Finkbeiner}]{Green2019}
Green, G.~M., Schla, E., Zucker, C., Speagle, J.~S., \& Finkbeiner, D. 2019,
  \apj, 93, \dodoi{10.3847/1538-4357/ab5362}

\bibitem[{{Greif}(2015)}]{Greif2015}
{Greif}, T.~H. 2015, Computational Astrophysics and Cosmology, 2, 3,
  \dodoi{10.1186/s40668-014-0006-2}

\bibitem[{{Greif} {et~al.}(2012){Greif}, {Bromm}, {Clark}, {Glover}, {Smith},
  {Klessen}, {Yoshida}, \& {Springel}}]{Greif2012}
{Greif}, T.~H., {Bromm}, V., {Clark}, P.~C., {et~al.} 2012, \mnras, 424, 399,
  \dodoi{10.1111/j.1365-2966.2012.21212.x}

\bibitem[{{Greif} {et~al.}(2011){Greif}, {Springel}, {White}, {Glover},
  {Clark}, {Smith}, {Klessen}, \& {Bromm}}]{Greif2011}
{Greif}, T.~H., {Springel}, V., {White}, S. D.~M., {et~al.} 2011, \apj, 737,
  75, \dodoi{10.1088/0004-637X/737/2/75}

\bibitem[{{Gunn} {et~al.}(1998){Gunn}, {Carr}, {Rockosi}, {Sekiguchi}, {Berry},
  {Elms}, {de Haas}, {Ivezi{\'c}}, {Knapp}, {Lupton}, {Pauls}, {Simcoe},
  {Hirsch}, {Sanford}, {Wang}, {York}, {Harris}, {Annis}, {Bartozek},
  {Boroski}, {Bakken}, {Haldeman}, {Kent}, {Holm}, {Holmgren}, {Petravick},
  {Prosapio}, {Rechenmacher}, {Doi}, {Fukugita}, {Shimasaku}, {Okada}, {Hull},
  {Siegmund}, {Mannery}, {Blouke}, {Heidtman}, {Schneider}, {Lucinio}, \&
  {Brinkman}}]{Gunn1998}
{Gunn}, J.~E., {Carr}, M., {Rockosi}, C., {et~al.} 1998, \aj, 116, 3040,
  \dodoi{10.1086/300645}

\bibitem[{{Gunn} {et~al.}(2006){Gunn}, {Siegmund}, {Mannery}, {Owen}, {Hull},
  {Leger}, {Carey}, {Knapp}, {York}, {Boroski}, {Kent}, {Lupton}, {Rockosi},
  {Evans}, {Waddell}, {Anderson}, {Annis}, {Barentine}, {Bartoszek}, {Bastian},
  {Bracker}, {Brewington}, {Briegel}, {Brinkmann}, {Brown}, {Carr},
  {Czarapata}, {Drennan}, {Dombeck}, {Federwitz}, {Gillespie}, {Gonzales},
  {Hansen}, {Harvanek}, {Hayes}, {Jordan}, {Kinney}, {Klaene}, {Kleinman},
  {Kron}, {Kresinski}, {Lee}, {Limmongkol}, {Lindenmeyer}, {Long}, {Loomis},
  {McGehee}, {Mantsch}, {Neilsen}, {Neswold}, {Newman}, {Nitta}, {Peoples},
  {Pier}, {Prieto}, {Prosapio}, {Rivetta}, {Schneider}, {Snedden}, \&
  {Wang}}]{Gunn2006}
{Gunn}, J.~E., {Siegmund}, W.~A., {Mannery}, E.~J., {et~al.} 2006, \aj, 131,
  2332, \dodoi{10.1086/500975}

\bibitem[{Hansen {et~al.}(2020)Hansen, Koch, Mashonkina, Magg, Bergemann,
  Sitnova, Gallagher, Ilyin, Caffau, Zhang, Strassmeier, \&
  Klessen}]{Hansen2020}
Hansen, C.~J., Koch, A., Mashonkina, L., {et~al.} 2020, Astronomy {\&}
  Astrophysics.
\newblock \doarXiv{2009.11876}

\bibitem[{Harris {et~al.}(2020)Harris, Millman, van~der Walt, Gommers,
  Virtanen, Cournapeau, Wieser, Taylor, Berg, Smith, Kern, Picus, Hoyer, van
  Kerkwijk, Brett, Haldane, del R{\'{i}}o, Wiebe, Peterson,
  G{\'{e}}rard-Marchant, Sheppard, Reddy, Weckesser, Abbasi, Gohlke, \&
  Oliphant}]{Harris2020}
Harris, C.~R., Millman, K.~J., van~der Walt, S.~J., {et~al.} 2020, Nature, 585,
  357, \dodoi{10.1038/s41586-020-2649-2}

\bibitem[{Hartwig {et~al.}(2015)Hartwig, Bromm, Klessen, \&
  Glover}]{Hartwig2015a}
Hartwig, T., Bromm, V., Klessen, R.~S., \& Glover, S.~C. 2015, Monthly Notices
  of the Royal Astronomical Society, 447, 3892, \dodoi{10.1093/mnras/stu2740}

\bibitem[{{Hirano} \& {Bromm}(2017)}]{Hirano2017}
{Hirano}, S., \& {Bromm}, V. 2017, \mnras, 470, 898,
  \dodoi{10.1093/mnras/stx1220}

\bibitem[{Hunter(2007)}]{Hunter2007}
Hunter, J.~D. 2007, Computing in Science and Engineering, 9, 90,
  \dodoi{10.1109/MCSE.2007.55}

\bibitem[{Husser {et~al.}(2013)Husser, {Wende-Von Berg}, Dreizler, Homeier,
  Reiners, Barman, \& Hauschildt}]{Husser2013}
Husser, T.~O., {Wende-Von Berg}, S., Dreizler, S., {et~al.} 2013, Astronomy and
  Astrophysics, 553, 1, \dodoi{10.1051/0004-6361/201219058}

\bibitem[{{Ishigaki} {et~al.}(2018){Ishigaki}, {Tominaga}, {Kobayashi}, \&
  {Nomoto}}]{Ishigaki2018}
{Ishigaki}, M.~N., {Tominaga}, N., {Kobayashi}, C., \& {Nomoto}, K. 2018, ApJ,
  857, 46, \dodoi{10.3847/1538-4357/aab3de}

\bibitem[{Ivezic {et~al.}(2019)Ivezic, Kahn, Tyson, Abel, Acosta, Allsman,
  Alonso, AlSayyad, Anderson, Andrew, {P. Angel}, Angeli, Ansari, Antilogus,
  Araujo, Armstrong, Arndt, Astier, Aubourg, Auza, Axelrod, Bard, Barr, Barrau,
  Bartlett, Bauer, Bauman, Baumont, Bechtol, Bechtol, Becker, Becla, Beldica,
  Bellavia, Bianco, Biswas, Blanc, Blazek, Blandford, Bloom, Bogart, Bond,
  Booth, Borgland, Borne, Bosch, Boutigny, Brackett, Bradshaw, Brandt, Brown,
  Bullock, Burchat, Burke, Cagnoli, Calabrese, Callahan, Callen, Carlin,
  Carlson, Chandrasekharan, Charles-Emerson, Chesley, Cheu, Chiang, Chiang,
  Chirino, Chow, Ciardi, Claver, Cohen-Tanugi, Cockrum, Coles, Connolly, Cook,
  Cooray, Covey, Cribbs, Cui, Cutri, Daly, Daniel, Daruich, Daubard, Daues,
  Dawson, Delgado, Dellapenna, Peyster, de~Val-Borro, Digel, Doherty, Dubois,
  Dubois-Felsmann, Durech, Economou, Eifler, Eracleous, Emmons, Neto, Ferguson,
  Figueroa, Fisher-Levine, Focke, Foss, Frank, Freemon, Gangler, Gawiser,
  Geary, Gee, Geha, Gessner, Gibson, Gilmore, Glanzman, Glick, Goldina,
  Goldstein, Goodenow, Graham, Gressler, Gris, Guy, Guyonnet, Haller, Harris,
  Hascall, Haupt, Hernandez, Herrmann, Hileman, Hoblitt, Hodgson, Hogan,
  Howard, Huang, Huffer, Ingraham, Innes, Jacoby, Jain, Jammes, Jee, Jenness,
  Jernigan, Jevremovi{\'{c}}, Johns, Johnson, Johnson, Jones, Juramy-Gilles,
  Juri{\'{c}}, Kalirai, Kallivayalil, Kalmbach, Kantor, Karst, Kasliwal, Kelly,
  Kessler, Kinnison, Kirkby, Knox, Kotov, Krabbendam, Krughoff, Kub{\'{a}}nek,
  Kuczewski, Kulkarni, Ku, Kurita, Lage, Lambert, Lange, Langton, Guillou,
  Levine, Liang, Lim, Lintott, Long, Lopez, Lotz, Lupton, Lust, MacArthur,
  Mahabal, Mandelbaum, Markiewicz, Marsh, Marshall, Marshall, May, McKercher,
  McQueen, Meyers, Migliore, Miller, Mills, Miraval, Moeyens, Moolekamp, Monet,
  Moniez, Monkewitz, Montgomery, Morrison, Mueller, Muller, Arancibia, Neill,
  Newbry, Nief, Nomerotski, Nordby, O'Connor, Oliver, Olivier, Olsen,
  O'Mullane, Ortiz, Osier, Owen, Pain, Palecek, Parejko, Parsons, Pease,
  Peterson, Peterson, Petravick, Petrick, Petry, Pierfederici, Pietrowicz,
  Pike, Pinto, Plante, Plate, Plutchak, Price, Prouza, Radeka, Rajagopal,
  Rasmussen, Regnault, Reil, Reiss, Reuter, Ridgway, Riot, Ritz, Robinson,
  Roby, Roodman, Rosing, Roucelle, Rumore, Russo, Saha, Sassolas, Schalk,
  Schellart, Schindler, Schmidt, Schneider, Schneider, Schoening, Schumacher,
  Schwamb, Sebag, Selvy, Sembroski, Seppala, Serio, Serrano, Shaw, Shipsey,
  Sick, Silvestri, Slater, Smith, Smith, Sobhani, Soldahl, Storrie-Lombardi,
  Stover, Strauss, Street, Stubbs, Sullivan, Sweeney, Swinbank, Szalay, Takacs,
  Tether, Thaler, Thayer, Thomas, Thornton, Thukral, Tice, Trilling, Turri,
  Berg, Berk, Vetter, Virieux, Vucina, Wahl, Walkowicz, Walsh, Walter, Wang,
  Wang, Warner, Wiecha, Willman, Winters, Wittman, Wolff, Wood-Vasey, Wu, Xin,
  Yoachim, \& Zhan}]{LSST2019}
Ivezic, Z., Kahn, S.~M., Tyson, J.~A., {et~al.} 2019, The Astrophysical
  Journal, 873, 111, \dodoi{10.3847/1538-4357/ab042c}

\bibitem[{Kepler {et~al.}(2019)Kepler, Pelisoli, Koester, Reindl, Geier,
  Romero, Ourique, {De Paula Oliveira}, \& Amaral}]{Kepler2019a}
Kepler, S.~O., Pelisoli, I., Koester, D., {et~al.} 2019, Monthly Notices of the
  Royal Astronomical Society, 486, 2169, \dodoi{10.1093/mnras/stz960}

\bibitem[{{Kobayashi} {et~al.}(2020){Kobayashi}, {Karakas}, \&
  {Lugaro}}]{Kobayashi2020}
{Kobayashi}, C., {Karakas}, A.~I., \& {Lugaro}, M. 2020, ApJ, 900, 179,
  \dodoi{10.3847/1538-4357/abae65}

\bibitem[{Koester {et~al.}(2014)Koester, G{\"{a}}nsicke, \&
  Farihi}]{Koester2014}
Koester, D., G{\"{a}}nsicke, B.~T., \& Farihi, J. 2014, Astronomy and
  Astrophysics, 566, \dodoi{10.1051/0004-6361/201423691}

\bibitem[{Kollmeier {et~al.}(2017)Kollmeier, Zasowski, Rix, Johns, Anderson,
  Drory, Johnson, Pogge, Bird, Blanc, Brownstein, Crane, {De Lee}, Klaene,
  Kreckel, MacDonald, Merloni, Ness, O'Brien, Sanchez-Gallego, Sayres, Shen,
  Thakar, Tkachenko, Aerts, Blanton, Eisenstein, Holtzman, Maoz, Nandra,
  Rockosi, Weinberg, Bovy, Casey, Chaname, Clerc, Conroy, Eracleous,
  G{\"{a}}nsicke, Hekker, Horne, Kauffmann, McQuinn, Pellegrini, Schinnerer,
  Schlafly, Schwope, Seibert, Teske, \& van Saders}]{Kollmeier2017}
Kollmeier, J.~A., Zasowski, G., Rix, H.-W., {et~al.} 2017, in Bulletin of the
  American Astronomical Society, 274.
\newblock \doarXiv{1711.03234}

\bibitem[{Kosakowski {et~al.}(2020)Kosakowski, Kilic, Brown, \&
  Gianninas}]{Kosakowski2020}
Kosakowski, A., Kilic, M., Brown, W.~R., \& Gianninas, A. 2020, The
  Astrophysical Journal, 894, 53, \dodoi{10.3847/1538-4357/ab8300}

\bibitem[{Kroupa(2001)}]{Kroupa2001}
Kroupa, P. 2001, Monthly Notices of the Royal Astronomical Society, 322, 231,
  \dodoi{10.1046/j.1365-8711.2001.04022.x}

\bibitem[{Kroupa(2002)}]{Kroupa2002a}
---. 2002, Science, 295, 82, \dodoi{10.1126/science.1067524}

\bibitem[{Lindegren {et~al.}(2018)Lindegren, Hern{\'{a}}ndez, Bombrun, Klioner,
  Bastian, Ramos-Lerate, {De Torres}, Steidelm{\"{o}}ller, Stephenson, Hobbs,
  Lammers, Biermann, Geyer, Hilger, Michalik, Stampa, McMillan,
  Casta{\~{n}}eda, Clotet, Comoretto, Davidson, Fabricius, Gracia, Hambly,
  Hutton, Mora, Portell, {Van Leeuwen}, Abbas, Abreu, Altmann, Andrei, Anglada,
  Balaguer-N{\'{u}}{\~{n}}ez, Barache, Becciani, Bertone, Bianchi, Bouquillon,
  Bourda, Br{\"{o}}semeister, Bucciarelli, Busonero, Buzzi, Cancelliere,
  Carlucci, Charlot, Cheek, Crosta, Crowley, {De Bruijne}, {De Felice},
  Drimmel, Esquej, Fienga, Fraile, Gai, Garralda, Gonz{\'{a}}lez-Vidal, Guerra,
  Hauser, Hofmann, Holl, Jordan, Lattanzi, Lenhardt, Liao, Licata, Lister,
  L{\"{o}}ffler, Marchant, Martin-Fleitas, Messineo, Mignard, Morbidelli,
  Poggio, Riva, Rowell, Salguero, Sarasso, Sciacca, Siddiqui, Smart, Spagna,
  Steele, Taris, Torra, {Van Elteren}, {Van Reeven}, \&
  Vecchiato}]{Lindegren2018}
Lindegren, L., Hern{\'{a}}ndez, J., Bombrun, A., {et~al.} 2018, Astronomy and
  Astrophysics, 616, \dodoi{10.1051/0004-6361/201832727}

\bibitem[{{Lokhorst} {et~al.}(2016){Lokhorst}, {Starkenburg}, {McConnachie},
  {Navarro}, {Ferrarese}, {C{\^o}t{\'e}}, {Liu}, {Peng}, {Gwyn}, {Cuillandre},
  \& {Guhathakurta}}]{Lokhorst2016}
{Lokhorst}, D., {Starkenburg}, E., {McConnachie}, A.~W., {et~al.} 2016, ApJ,
  819, 124, \dodoi{10.3847/0004-637X/819/2/124}

\bibitem[{Luri {et~al.}(2018)Luri, Brown, Sarro, Arenou, Bailer-Jones,
  Castro-Ginard, Bruijne, Prusti, Babusiaux, \& Delgado}]{Luri2018}
Luri, X., Brown, A.~G., Sarro, L.~M., {et~al.} 2018, Astronomy and
  Astrophysics, 616, A9, \dodoi{10.1051/0004-6361/201832964}

\bibitem[{{Luyten}(1922)}]{Luyten1922}
{Luyten}, W.~J. 1922, LicOB, 336, 135

\bibitem[{Magg {et~al.}(2019)Magg, Klessen, Glover, \& Li}]{Magg2019}
Magg, M., Klessen, R.~S., Glover, S.~C., \& Li, H. 2019, Monthly Notices of the
  Royal Astronomical Society, 487, 486, \dodoi{10.1093/mnras/stz1210}

\bibitem[{{Marrese} {et~al.}(2019){Marrese}, {Marinoni}, {Fabrizio}, \&
  {Altavilla}}]{Marrese2019}
{Marrese}, P.~M., {Marinoni}, S., {Fabrizio}, M., \& {Altavilla}, G. 2019,
  \aap, 621, A144, \dodoi{10.1051/0004-6361/201834142}

\bibitem[{Marshall {et~al.}(2006)Marshall, Robin, Reyl{\'{e}}, Schultheis, \&
  Picaud}]{Marshall2006}
Marshall, D.~J., Robin, A.~C., Reyl{\'{e}}, C., Schultheis, M., \& Picaud, S.
  2006, Astronomy {\&} Astrophysics, 651, 635

\bibitem[{Newville {et~al.}(2014)Newville, Ingargiola, Stensitzki, \&
  Allen}]{Newville2014}
Newville, M., Ingargiola, A., Stensitzki, T., \& Allen, D.~B. 2014, Zenodo, ,
  \dodoi{10.5281/ZENODO.11813}

\bibitem[{Niculescu-Mizil \& Caruana(2005)}]{Niculescu2005}
Niculescu-Mizil, A., \& Caruana, R. 2005, in Proceedings of the 22nd
  International Conference on Machine Learning, ICML '05 (New York, NY, USA:
  Association for Computing Machinery), 625–632,
  \dodoi{10.1145/1102351.1102430}

\bibitem[{{Padmanabhan} {et~al.}(2008){Padmanabhan}, {Schlegel}, {Finkbeiner},
  {Barentine}, {Blanton}, {Brewington}, {Gunn}, {Harvanek}, {Hogg},
  {Ivezi{\'c}}, {Johnston}, {Kent}, {Kleinman}, {Knapp}, {Krzesinski}, {Long},
  {Neilsen}, {Nitta}, {Loomis}, {Lupton}, {Roweis}, {Snedden}, {Strauss}, \&
  {Tucker}}]{Padmanabhan2008}
{Padmanabhan}, N., {Schlegel}, D.~J., {Finkbeiner}, D.~P., {et~al.} 2008, \apj,
  674, 1217, \dodoi{10.1086/524677}

\bibitem[{{Paxton} {et~al.}(2011){Paxton}, {Bildsten}, {Dotter}, {Herwig},
  {Lesaffre}, \& {Timmes}}]{Paxton2011}
{Paxton}, B., {Bildsten}, L., {Dotter}, A., {et~al.} 2011, ApJS, 192, 3,
  \dodoi{10.1088/0067-0049/192/1/3}

\bibitem[{{Paxton} {et~al.}(2013){Paxton}, {Cantiello}, {Arras}, {Bildsten},
  {Brown}, {Dotter}, {Mankovich}, {Montgomery}, {Stello}, {Timmes}, \&
  {Townsend}}]{Paxton2013}
{Paxton}, B., {Cantiello}, M., {Arras}, P., {et~al.} 2013, ApJS, 208, 4,
  \dodoi{10.1088/0067-0049/208/1/4}

\bibitem[{{Paxton} {et~al.}(2015){Paxton}, {Marchant}, {Schwab}, {Bauer},
  {Bildsten}, {Cantiello}, {Dessart}, {Farmer}, {Hu}, {Langer}, {Townsend},
  {Townsley}, \& {Timmes}}]{Paxton2015}
{Paxton}, B., {Marchant}, P., {Schwab}, J., {et~al.} 2015, ApJS, 220, 15,
  \dodoi{10.1088/0067-0049/220/1/15}

\bibitem[{{Paxton} {et~al.}(2018){Paxton}, {Schwab}, {Bauer}, {Bildsten},
  {Blinnikov}, {Duffell}, {Farmer}, {Goldberg}, {Marchant}, {Sorokina},
  {Thoul}, {Townsend}, \& {Timmes}}]{Paxton2018}
{Paxton}, B., {Schwab}, J., {Bauer}, E.~B., {et~al.} 2018, ApJS, 234, 34,
  \dodoi{10.3847/1538-4365/aaa5a8}

\bibitem[{Pedregosa {et~al.}(2011)Pedregosa, Michel, Grisel, Blondel,
  Prettenhofer, Weiss, Vanderplas, Cournapeau, Pedregosa, Varoquaux, Gramfort,
  Thirion, Grisel, Dubourg, Passos, \& Brucher}]{Pedregosa2011}
Pedregosa, F., Michel, V., Grisel, O., {et~al.} 2011, Journal of Machine
  Learning Research, 12, 2825.
\newblock \url{http://scikit-learn.sourceforge.net.}

\bibitem[{{Pelisoli} {et~al.}(2019){Pelisoli}, {Bell}, {Kepler}, \&
  {Koester}}]{Pelisoli2019b}
{Pelisoli}, I., {Bell}, K.~J., {Kepler}, S.~O., \& {Koester}, D. 2019, MNRAS,
  482, 3831, \dodoi{10.1093/mnras/sty2979}

\bibitem[{Pelisoli {et~al.}(2018{\natexlab{a}})Pelisoli, Kepler, \&
  Koester}]{Pelisoli2018b}
Pelisoli, I., Kepler, S.~O., \& Koester, D. 2018{\natexlab{a}}, Monthly Notices
  of the Royal Astronomical Society, 475, 2480, \dodoi{10.1093/mnras/sty011}

\bibitem[{Pelisoli {et~al.}(2018{\natexlab{b}})Pelisoli, Kepler, Koester,
  Castanheira, Romero, \& Fraga}]{Pelisoli2018}
Pelisoli, I., Kepler, S.~O., Koester, D., {et~al.} 2018{\natexlab{b}}, Monthly
  Notices of the Royal Astronomical Society, 478, 867,
  \dodoi{10.1093/MNRAS/STY1101}

\bibitem[{{Pelisoli} {et~al.}(2017){Pelisoli}, {Kepler}, {Koester}, \&
  {Romero}}]{Pelisoli2016}
{Pelisoli}, I., {Kepler}, S.~O., {Koester}, D., \& {Romero}, A.~D. 2017, in
  Astronomical Society of the Pacific Conference Series, Vol. 509, 20th
  European White Dwarf Workshop, ed. P.~E. {Tremblay}, B.~{Gaensicke}, \&
  T.~{Marsh}, 447.
\newblock \doarXiv{1610.05550}

\bibitem[{{Pelisoli} \& {Vos}(2019)}]{Pelisoli2019a}
{Pelisoli}, I., \& {Vos}, J. 2019, MNRAS, 488, 2892,
  \dodoi{10.1093/mnras/stz1876}

\bibitem[{{Placco} {et~al.}(2015){Placco}, {Frebel}, {Lee}, {Jacobson},
  {Beers}, {Pena}, {Chan}, \& {Heger}}]{Placco2015}
{Placco}, V.~M., {Frebel}, A., {Lee}, Y.~S., {et~al.} 2015, \apj, 809, 136,
  \dodoi{10.1088/0004-637X/809/2/136}

\bibitem[{{Placco} {et~al.}(2016){Placco}, {Frebel}, {Beers}, {Yoon}, {Chiti},
  {Heger}, {Chan}, {Casey}, \& {Christlieb}}]{Placco2016}
{Placco}, V.~M., {Frebel}, A., {Beers}, T.~C., {et~al.} 2016, \apj, 833, 21,
  \dodoi{10.3847/0004-637X/833/1/21}

\bibitem[{{Riaz} {et~al.}(2018){Riaz}, {Bovino}, {Vanaverbeke}, \&
  {Schleicher}}]{Riaz2018}
{Riaz}, R., {Bovino}, S., {Vanaverbeke}, S., \& {Schleicher}, D.~R.~G. 2018,
  \mnras, 479, 667, \dodoi{10.1093/mnras/sty1635}

\bibitem[{Robitaille {et~al.}(2013)Robitaille, Tollerud, Greenfield,
  Droettboom, Bray, Aldcroft, Davis, Ginsburg, Price-Whelan, Kerzendorf,
  Conley, Crighton, Barbary, Muna, Ferguson, Grollier, Parikh, Nair,
  G{\"{u}}nther, Deil, Woillez, Conseil, Kramer, Turner, Singer, Fox, Weaver,
  Zabalza, Edwards, {Azalee Bostroem}, Burke, Casey, Crawford, Dencheva, Ely,
  Jenness, Labrie, Lim, Pierfederici, Pontzen, Ptak, Refsdal, Servillat, \&
  Streicher}]{Robitaille2013}
Robitaille, T.~P., Tollerud, E.~J., Greenfield, P., {et~al.} 2013, Astronomy
  and Astrophysics, 558, 1, \dodoi{10.1051/0004-6361/201322068}

\bibitem[{{Roederer} {et~al.}(2014){Roederer}, {Preston}, {Thompson},
  {Shectman}, {Sneden}, {Burley}, \& {Kelson}}]{2014AJ....147..136R}
{Roederer}, I.~U., {Preston}, G.~W., {Thompson}, I.~B., {et~al.} 2014, \aj,
  147, 136, \dodoi{10.1088/0004-6256/147/6/136}

\bibitem[{{Ryan} \& {Norris}(1991{\natexlab{a}})}]{1991AJ....101.1835R}
{Ryan}, S.~G., \& {Norris}, J.~E. 1991{\natexlab{a}}, \aj, 101, 1835,
  \dodoi{10.1086/115811}

\bibitem[{{Ryan} \& {Norris}(1991{\natexlab{b}})}]{1991AJ....101.1865R}
---. 1991{\natexlab{b}}, \aj, 101, 1865, \dodoi{10.1086/115812}

\bibitem[{{Ryan} {et~al.}(1991){Ryan}, {Norris}, \&
  {Bessell}}]{1991AJ....102..303R}
{Ryan}, S.~G., {Norris}, J.~E., \& {Bessell}, M.~S. 1991, \aj, 102, 303,
  \dodoi{10.1086/115878}

\bibitem[{{Salgado} {et~al.}(2017){Salgado}, {Gonz{\'a}lez-N{\'u}{\~n}ez},
  {Guti{\'e}rrez-S{\'a}nchez}, {Segovia}, {Dur{\'a}n}, {Hern{\'a}ndez}, \&
  {Arviset}}]{Salgado2017}
{Salgado}, J., {Gonz{\'a}lez-N{\'u}{\~n}ez}, J., {Guti{\'e}rrez-S{\'a}nchez},
  R., {et~al.} 2017, Astronomy and Computing, 21, 22,
  \dodoi{10.1016/j.ascom.2017.08.002}

\bibitem[{{Saumon} {et~al.}(1994){Saumon}, {Bergeron}, {Lunine}, {Hubbard}, \&
  {Burrows}}]{Saumon1994}
{Saumon}, D., {Bergeron}, P., {Lunine}, J.~I., {Hubbard}, W.~B., \& {Burrows},
  A. 1994, \apj, 424, 333, \dodoi{10.1086/173892}

\bibitem[{{Schatzman}(1948)}]{Schatzman1948}
{Schatzman}, E. 1948, \nat, 161, 61, \dodoi{10.1038/161061b0}

\bibitem[{Schlaufman {et~al.}(2018)Schlaufman, Thompson, \&
  Casey}]{Schlaufman2018}
Schlaufman, K.~C., Thompson, I.~B., \& Casey, A.~R. 2018, The Astrophysical
  Journal, 867, 98, \dodoi{10.3847/1538-4357/aadd97}

\bibitem[{{Silk}(1983)}]{Silk1983}
{Silk}, J. 1983, \mnras, 205, 705, \dodoi{10.1093/mnras/205.3.705}

\bibitem[{{Smee} {et~al.}(2013){Smee}, {Gunn}, {Uomoto}, {Roe}, {Schlegel},
  {Rockosi}, {Carr}, {Leger}, {Dawson}, {Olmstead}, {Brinkmann}, {Owen},
  {Barkhouser}, {Honscheid}, {Harding}, {Long}, {Lupton}, {Loomis}, {Anderson},
  {Annis}, {Bernardi}, {Bhardwaj}, {Bizyaev}, {Bolton}, {Brewington}, {Briggs},
  {Burles}, {Burns}, {Castander}, {Connolly}, {Davenport}, {Ebelke}, {Epps},
  {Feldman}, {Friedman}, {Frieman}, {Heckman}, {Hull}, {Knapp}, {Lawrence},
  {Loveday}, {Mannery}, {Malanushenko}, {Malanushenko}, {Merrelli}, {Muna},
  {Newman}, {Nichol}, {Oravetz}, {Pan}, {Pope}, {Ricketts}, {Shelden},
  {Sandford}, {Siegmund}, {Simmons}, {Smith}, {Snedden}, {Schneider},
  {SubbaRao}, {Tremonti}, {Waddell}, \& {York}}]{Smee2013}
{Smee}, S.~A., {Gunn}, J.~E., {Uomoto}, A., {et~al.} 2013, \aj, 146, 32,
  \dodoi{10.1088/0004-6256/146/2/32}

\bibitem[{{Smith} {et~al.}(2002){Smith}, {Tucker}, {Kent}, {Richmond},
  {Fukugita}, {Ichikawa}, {Ichikawa}, {Jorgensen}, {Uomoto}, {Gunn}, {Hamabe},
  {Watanabe}, {Tolea}, {Henden}, {Annis}, {Pier}, {McKay}, {Brinkmann}, {Chen},
  {Holtzman}, {Shimasaku}, \& {York}}]{Smith2002}
{Smith}, J.~A., {Tucker}, D.~L., {Kent}, S., {et~al.} 2002, \aj, 123, 2121,
  \dodoi{10.1086/339311}

\bibitem[{{Stacy} \& {Bromm}(2013)}]{Stacy2013}
{Stacy}, A., \& {Bromm}, V. 2013, \mnras, 433, 1094,
  \dodoi{10.1093/mnras/stt789}

\bibitem[{{Stacy} \& {Bromm}(2014)}]{Stacy2014}
---. 2014, \apj, 785, 73, \dodoi{10.1088/0004-637X/785/1/73}

\bibitem[{{Stacy} {et~al.}(2016){Stacy}, {Bromm}, \& {Lee}}]{Stacy2016}
{Stacy}, A., {Bromm}, V., \& {Lee}, A.~T. 2016, \mnras, 462, 1307,
  \dodoi{10.1093/mnras/stw1728}

\bibitem[{{Stacy} {et~al.}(2010){Stacy}, {Greif}, \& {Bromm}}]{Stacy2010}
{Stacy}, A., {Greif}, T.~H., \& {Bromm}, V. 2010, \mnras, 403, 45,
  \dodoi{10.1111/j.1365-2966.2009.16113.x}

\bibitem[{{Stacy} {et~al.}(2012){Stacy}, {Greif}, \& {Bromm}}]{Stacy2012}
---. 2012, \mnras, 422, 290, \dodoi{10.1111/j.1365-2966.2012.20605.x}

\bibitem[{{Suda} {et~al.}(2011){Suda}, {Yamada}, {Katsuta}, {Komiya},
  {Ishizuka}, {Aoki}, \& {Fujimoto}}]{Suda2011}
{Suda}, T., {Yamada}, S., {Katsuta}, Y., {et~al.} 2011, \mnras, 412, 843,
  \dodoi{10.1111/j.1365-2966.2011.17943.x}

\bibitem[{{Suda} {et~al.}(2008){Suda}, {Katsuta}, {Yamada}, {Suwa}, {Ishizuka},
  {Komiya}, {Sorai}, {Aikawa}, \& {Fujimoto}}]{Suda2008}
{Suda}, T., {Katsuta}, Y., {Yamada}, S., {et~al.} 2008, \pasj, 60, 1159,
  \dodoi{10.1093/pasj/60.5.1159}

\bibitem[{{Suda} {et~al.}(2017){Suda}, {Hidaka}, {Aoki}, {Katsuta}, {Yamada},
  {Fujimoto}, {Ohtani}, {Masuyama}, {Noda}, \& {Wada}}]{Suda2017}
{Suda}, T., {Hidaka}, J., {Aoki}, W., {et~al.} 2017, \pasj, 69, 76,
  \dodoi{10.1093/pasj/psx059}

\bibitem[{{Tegmark} {et~al.}(1997){Tegmark}, {Silk}, {Rees}, {Blanchard},
  {Abel}, \& {Palla}}]{Tegmark1997}
{Tegmark}, M., {Silk}, J., {Rees}, M.~J., {et~al.} 1997, \apj, 474, 1,
  \dodoi{10.1086/303434}

\bibitem[{Tremblay \& Bergeron(2009)}]{Tremblay2009}
Tremblay, P.~E., \& Bergeron, P. 2009, Astrophysical Journal, 696, 1755,
  \dodoi{10.1088/0004-637X/696/2/1755}

\bibitem[{Virtanen {et~al.}(2020)Virtanen, Gommers, Oliphant, Haberland, Reddy,
  Cournapeau, Burovski, Peterson, Weckesser, Bright, van~der Walt, Brett,
  Wilson, Millman, Mayorov, Nelson, Jones, Kern, Larson, Carey, Polat, Feng,
  Moore, VanderPlas, Laxalde, Perktold, Cimrman, Henriksen, Quintero, Harris,
  Archibald, Ribeiro, Pedregosa, van Mulbregt, Vijaykumar, Bardelli, Rothberg,
  Hilboll, Kloeckner, Scopatz, Lee, Rokem, Woods, Fulton, Masson,
  H{\"{a}}ggstr{\"{o}}m, Fitzgerald, Nicholson, Hagen, Pasechnik, Olivetti,
  Martin, Wieser, Silva, Lenders, Wilhelm, Young, Price, Ingold, Allen, Lee,
  Audren, Probst, Dietrich, Silterra, Webber, Slavi{\v{c}}, Nothman, Buchner,
  Kulick, Sch{\"{o}}nberger, {de Miranda Cardoso}, Reimer, Harrington,
  Rodr{\'{i}}guez, Nunez-Iglesias, Kuczynski, Tritz, Thoma, Newville,
  K{\"{u}}mmerer, Bolingbroke, Tartre, Pak, Smith, Nowaczyk, Shebanov, Pavlyk,
  Brodtkorb, Lee, McGibbon, Feldbauer, Lewis, Tygier, Sievert, Vigna, Peterson,
  More, Pudlik, Oshima, Pingel, Robitaille, Spura, Jones, Cera, Leslie, Zito,
  Krauss, Upadhyay, Halchenko, \& V{\'{a}}zquez-Baeza}]{Virtanen2020}
Virtanen, P., Gommers, R., Oliphant, T.~E., {et~al.} 2020, Nature Methods, 17,
  261, \dodoi{10.1038/s41592-019-0686-2}

\bibitem[{{Wenger} {et~al.}(2000){Wenger}, {Ochsenbein}, {Egret}, {Dubois},
  {Bonnarel}, {Borde}, {Genova}, {Jasniewicz}, {Lalo{\"e}}, {Lesteven}, \&
  {Monier}}]{Wenger2000}
{Wenger}, M., {Ochsenbein}, F., {Egret}, D., {et~al.} 2000, \aaps, 143, 9,
  \dodoi{10.1051/aas:2000332}

\bibitem[{{Wollenberg} {et~al.}(2020){Wollenberg}, {Glover}, {Clark}, \&
  {Klessen}}]{Wollenberg2020}
{Wollenberg}, K. M.~J., {Glover}, S. C.~O., {Clark}, P.~C., \& {Klessen}, R.~S.
  2020, \mnras, 494, 1871, \dodoi{10.1093/mnras/staa289}

\bibitem[{{Yamada} {et~al.}(2013){Yamada}, {Suda}, {Komiya}, {Aoki}, \&
  {Fujimoto}}]{Yamada2013}
{Yamada}, S., {Suda}, T., {Komiya}, Y., {Aoki}, W., \& {Fujimoto}, M.~Y. 2013,
  \mnras, 436, 1362, \dodoi{10.1093/mnras/stt1652}

\bibitem[{{Yanny} {et~al.}(2009){Yanny}, {Rockosi}, {Newberg}, {Knapp},
  {Adelman-McCarthy}, {Alcorn}, {Allam}, {Allende Prieto}, {An}, {Anderson},
  {Anderson}, {Bailer-Jones}, {Bastian}, {Beers}, {Bell}, {Belokurov},
  {Bizyaev}, {Blythe}, {Bochanski}, {Boroski}, {Brinchmann}, {Brinkmann},
  {Brewington}, {Carey}, {Cudworth}, {Evans}, {Evans}, {Gates}, {G{\"a}nsicke},
  {Gillespie}, {Gilmore}, {Nebot Gomez-Moran}, {Grebel}, {Greenwell}, {Gunn},
  {Jordan}, {Jordan}, {Harding}, {Harris}, {Hendry}, {Holder}, {Ivans},
  {Ivezi{\v{c}}}, {Jester}, {Johnson}, {Kent}, {Kleinman}, {Kniazev},
  {Krzesinski}, {Kron}, {Kuropatkin}, {Lebedeva}, {Lee}, {French Leger},
  {L{\'e}pine}, {Levine}, {Lin}, {Long}, {Loomis}, {Lupton}, {Malanushenko},
  {Malanushenko}, {Margon}, {Martinez-Delgado}, {McGehee}, {Monet}, {Morrison},
  {Munn}, {Neilsen}, {Nitta}, {Norris}, {Oravetz}, {Owen}, {Padmanabhan},
  {Pan}, {Peterson}, {Pier}, {Platson}, {Re Fiorentin}, {Richards}, {Rix},
  {Schlegel}, {Schneider}, {Schreiber}, {Schwope}, {Sibley}, {Simmons},
  {Snedden}, {Allyn Smith}, {Stark}, {Stauffer}, {Steinmetz}, {Stoughton},
  {SubbaRao}, {Szalay}, {Szkody}, {Thakar}, {Sivarani}, {Tucker}, {Uomoto},
  {Vanden Berk}, {Vidrih}, {Wadadekar}, {Watters}, {Wilhelm}, {Wyse}, {Yarger},
  \& {Zucker}}]{Yanny2009}
{Yanny}, B., {Rockosi}, C., {Newberg}, H.~J., {et~al.} 2009, \aj, 137, 4377,
  \dodoi{10.1088/0004-6256/137/5/4377}

\bibitem[{{Yong} {et~al.}(2013){Yong}, {Norris}, {Bessell}, {Christlieb},
  {Asplund}, {Beers}, {Barklem}, {Frebel}, \& {Ryan}}]{2013ApJ...762...26Y}
{Yong}, D., {Norris}, J.~E., {Bessell}, M.~S., {et~al.} 2013, \apj, 762, 26,
  \dodoi{10.1088/0004-637X/762/1/26}

\bibitem[{Yoon {et~al.}(2016)Yoon, Beers, Placco, Rasmussen, Carollo, He,
  Hansen, Roederer, \& Zeanah}]{Yoon2016}
Yoon, J., Beers, T.~C., Placco, V.~M., {et~al.} 2016, The Astrophysical
  Journal, 833, 20, \dodoi{10.3847/0004-637x/833/1/20}

\bibitem[{York {et~al.}(2000)York, Adelman, {Anderson, Jr.}, Anderson, Annis,
  Bahcall, Bakken, Barkhouser, Bastian, Berman, Boroski, Bracker, Briegel,
  Briggs, Brinkmann, Brunner, Burles, Carey, Carr, Castander, Chen, Colestock,
  Connolly, Crocker, Csabai, Czarapata, Davis, Doi, Dombeck, Eisenstein,
  Ellman, Elms, Evans, Fan, Federwitz, Fiscelli, Friedman, Frieman, Fukugita,
  Gillespie, Gunn, Gurbani, de~Haas, Haldeman, Harris, Hayes, Heckman,
  Hennessy, Hindsley, Holm, Holmgren, Huang, Hull, Husby, Ichikawa, Ichikawa,
  Ivezi{\'{c}}, Kent, Kim, Kinney, Klaene, Kleinman, Kleinman, Knapp, Korienek,
  Kron, Kunszt, Lamb, Lee, Leger, Limmongkol, Lindenmeyer, Long, Loomis,
  Loveday, Lucinio, Lupton, MacKinnon, Mannery, Mantsch, Margon, McGehee,
  McKay, Meiksin, Merelli, Monet, Munn, Narayanan, Nash, Neilsen, Neswold,
  Newberg, Nichol, Nicinski, Nonino, Okada, Okamura, Ostriker, Owen, Pauls,
  Peoples, Peterson, Petravick, Pier, Pope, Pordes, Prosapio, Rechenmacher,
  Quinn, Richards, Richmond, Rivetta, Rockosi, Ruthmansdorfer, Sandford,
  Schlegel, Schneider, Sekiguchi, Sergey, Shimasaku, Siegmund, Smee, Smith,
  Snedden, Stone, Stoughton, Strauss, Stubbs, SubbaRao, Szalay, Szapudi,
  Szokoly, Thakar, Tremonti, Tucker, Uomoto, {Vanden Berk}, Vogeley, Waddell,
  Wang, Watanabe, Weinberg, Yanny, \& Yasuda}]{York2000}
York, D.~G., Adelman, J., {Anderson, Jr.}, J.~E., {et~al.} 2000, The
  Astronomical Journal, \dodoi{10.1086/301513}

\bibitem[{Yu {et~al.}(2011)Yu, Huang, \& Lin}]{Yu2011}
Yu, H.-F., Huang, F.-L., \& Lin, C.-J. 2011, Machine Learning, 85, 41,
  \dodoi{10.1007/s10994-010-5221-8}

\end{thebibliography}
\bibliographystyle{aasjournal}

% \listofchanges

\end{document}